\documentclass[acmlarge,nonacm]{acmart}
\AtBeginDocument{%
  }

\usepackage{todonotes}
\usepackage{booktabs}
\usepackage{multirow}
\usepackage{threeparttable}
\usepackage{tcolorbox}
\usepackage{tabularx}
\usepackage{ifthen}
\newboolean{showcomments}
\setboolean{showcomments}{true} 
\ifthenelse{\boolean{showcomments}}
{\newcommand{\nb}[2]{
        \fcolorbox{black}{yellow}{\bfseries\sffamily\scriptsize#1}
        {\sffamily\small$\blacktriangleright$#2$\blacktriangleleft$}
    }
}
{\newcommand{\nb}[2]{}
}

\setcopyright{acmlicensed}
\copyrightyear{2026}
\acmYear{2026}

\acmJournal{TOSEM}

\begin{document}

\title{Augmentation with Dilution? A Large-Scale Empirical Study of Human Contributor Ecosystems After AI Coding Agent Adoption}

\author{Weixing Zhang}
\email{weixing.zhang@kit.edu}
\orcid{0000-0003-2890-6034}
\correspondingauthor
\affiliation{%
  \institution{Karlsruhe Institute of Technology}
  \city{Karlsruhe}
  \country{Germany}
}

\author{Bowen Jiang}
\affiliation{%
  \institution{Karlsruhe Institute of Technology}
  \city{Karlsruhe}
  \country{Germany}}
\email{bowen.jiang@kit.edu}

\author{Tianhai Liu}
\affiliation{%
  \institution{Karlsruhe Institute of Technology}
  \city{Karlsruhe}
  \country{Germany}
}
\email{tianhai.liu@kit.edu}


\author{Jialong Li}
\affiliation{%
  \institution{Waseda University}
  \city{Tokyo}
  \country{Japan}}
\email{lijialong@fuji.waseda.jp}

\author{Anne Koziolek}
\affiliation{%
  \institution{Karlsruhe Institute of Technology}
  \city{Karlsruhe}
  \country{Germany}}
\email{koziolek@kit.edu}

\renewcommand{\shortauthors}{Zhang et al.}

\begin{abstract}
AI coding agents are rapidly proliferating in open-source software development, and prior research has established that the long-term sustainability of open-source projects is closely tied to the health of their human contributor ecosystems, including a steady influx of newcomers, the retention of core developers, and a diverse contributor base. Yet most existing studies treat human contributors as a static backdrop rather than as the subject of inquiry, and the available evidence on human participation after agent adoption mainly concerns newcomer outcomes. This paper takes the human contributor ecosystem as its dependent variable. To the best of our knowledge, it is the first study to estimate how agent adoption changes the number of distinct humans behind each pull request, and to relate this change to agent pull requests recorded under developer accounts. We identify treated repositories using the AIDev dataset combined with GitHub API scanning, apply agent-trace screening and two-stage propensity score matching, and construct a balanced panel of 2,246 treated and 3,798 control repositories spanning January 2023 to May 2026. Effects are estimated using the Sun--Abraham interaction-weighted estimator for staggered difference-in-differences, complemented by event study plots and HonestDiD sensitivity analysis to assess the parallel trends assumption. We find that AI agent adoption is associated with a decline in human contributor density of 0.019 to 0.048, about 4\%--10\% of its pre-adoption level, depending on the reference month (both $p < 0.01$); because this estimate tolerates only small violations of parallel trends, we regard it as suggestive evidence. The relative participation share of newcomers, however, shows no detectable change (ATT $= -0.009$, $p = 0.28$), with the 95\% confidence interval ruling out a decrease of more than 2.4 percentage points. Whether the decline in density is associated with the share of agent pull requests recorded under developer accounts remains inconclusive (the difference between repositories with high and low shares: $-0.013$, 95\% CI $[-0.083, 0.057]$). 
These findings are consistent with a pattern of \emph{augmentation with dilution}: we detect no decrease in the relative participation share of newcomers, but each human contributor appears to correspond to a larger number of pull requests.
\end{abstract}




\maketitle

\section{Introduction}
\label{sec:intro}
AI coding agents have entered open-source software development at an unprecedented pace. Robbes et al.~\cite{robbes2026agentic} estimate that between 22\% and 29\% of GitHub projects had adopted coding agents as of February 2026. At the commit level, the median number of lines added in agent-assisted commits is about three times that of human-authored commits, and agent-assisted commits consist predominantly of feature additions and bug fixes. GitHub reports that the share of pull requests started by an agent on its platform rose from 17\% in March 2026 to more than 40\% in September 2026; because only agents that GitHub can detect are counted, this figure should be regarded as a lower bound~\cite{github2026octoverse}. At the same time, a large body of empirical research has shown that the long-term development of open-source projects depends heavily on the health of their human contributor ecosystems: a steady influx of newcomers~\cite{steinmacher2015systematic}, the retention of established contributors~\cite{miller2019people}, and gender and tenure diversity among contributors~\cite{vasilescu2015gender} have each been linked to the continuity or productivity of open-source projects.

To date, empirical research on AI coding agents has focused mainly on what agents produce. Prior work has characterized the structural properties of AI-generated pull requests~\cite{ogenrwot2026ai}, measured the survival and modification patterns of AI-authored code~\cite{rahman2026will}, and quantified the technical debt it introduces~\cite{liu2026debt}. Other studies have examined how AI-generated pull requests are reviewed and merged, finding that a substantial share are merged with little or no explicit review~\cite{gao2026autopilot, watanabe2026use}, and have estimated the effects of agent adoption on development velocity and software quality at the project level~\cite{agarwal2026agents, he2026cursor}. Far less is known about the human side: what happens to the developers who used to write that code? The closest evidence concerns newcomers. Xu et al.~\cite{xu2026decoupling} date adoption by the first commit of an agent configuration file and find that, in established projects, adoption does not reduce newcomer inflow, the handling of first contributions, or newcomer retention. How adoption changes the overall structure of human participation, that is, how many distinct people stand behind a project’s pull request activity, has not been estimated.

Newcomers nevertheless remain a natural concern. Contributions by AI coding agents consist predominantly of features and fixes~\cite{robbes2026agentic}, although the observed task profile also depends on how agents are deployed and detected~\cite{khosravani2026detecting}. One of the main barriers newcomers face when joining open-source projects is finding a suitable task to start with~\cite{steinmacher2015systematic}. If agents systematically take over tasks that could otherwise serve as starting points, the contribution opportunities actually available to newcomers may shrink even as the absolute number of human contributors remains stable. There is already suggestive evidence: over 67.5\% of AI-co-authored pull requests originate from contributors without prior code ownership~\cite{gao2026autopilot}.

This paper addresses these gaps. We take the human contributor ecosystem as the dependent variable and examine how the participation structure of pull request activity and the relative share of newcomers change after AI coding agent adoption in open-source projects. We identify adopting repositories in the AIDev dataset~\cite{li2026aidev} and construct the control group through two-stage propensity score matching on pre-adoption levels and trends. Because GHArchive records a declining share of pull requests from mid-2025 onward, we retrieved the complete pull request history of each repository through the GitHub API. We use these data to compute all outcome variables, screen candidate control repositories for traces of agent use, correct adoption timing, and measure the share of agent pull requests recorded under developer accounts. We apply a staggered difference-in-differences design~\cite{callaway2021difference} with the Sun–Abraham estimator~\cite{sun2021estimating} to 2,246 treated and 3,798 control GitHub repositories observed from January 2023 to May 2026. With this design, we estimate changes in participation structure and newcomer relative share, and examine their association with agent pull requests recorded under developer accounts. To the best of our knowledge, this is the first study to estimate how agent adoption changes the number of distinct humans behind each pull request, and to relate this change to agent pull requests recorded under developer accounts.

Specifically, we address the following research questions:

\textbf{RQ1: How does the adoption of AI coding agents relate to changes in human contributor participation in open-source projects?}
A decline in human contributor density would indicate that fewer distinct humans stand behind each pull request, regardless of whether the absolute number of participants changes.

\textbf{RQ2: Does AI agent adoption reduce the relative participation share of newcomer contributors?}
Newcomers are of particular interest because the feature and fix tasks that agents take on may be precisely those that newcomers could otherwise use as starting points to build a contribution record.

\textbf{RQ3: Is the decrease in human contributor density associated with the share of agent pull requests recorded under developer accounts?}
A decrease in the number of distinct humans behind each pull request may reflect less human participation, or agent pull requests submitted under developers’ own accounts rather than under the agent’s bot account, which add to the count of pull requests but not to the count of people. Answering this question indicates whether part of an observed change in human participation reflects how agent work appears in repository records rather than a change in what humans do.

Our main findings are as follows. For RQ1, human contributor density declines by 0.019 to 0.048 after adoption, about 4\% to 10\% of its pre-adoption level depending on the reference month. The estimate is consistent across control group designs; Section~\ref{sec:humanpartici} reports its sensitivity to violations of parallel trends. For RQ2, we detect no change in the newcomer ratio (ATT $= -0.009$, $p = 0.28$), and the 95\% confidence interval rules out a decrease of more than 2.4 percentage points. For RQ3, the result is inconclusive: the difference in the density change between repositories with high and low shares of agent pull requests recorded under developer accounts is $-0.013$ (95\% CI $[-0.083, 0.057]$). These results are consistent with a pattern we call \emph{augmentation with dilution}: we find no evidence that agent adoption reduces the relative participation of newcomers, but each human contributor appears to correspond to more pull requests.

This paper makes the following contributions:
\begin{itemize}
  \item \textbf{Evidence on the structure of human participation.} We take the human contributor ecosystem as the dependent variable and estimate how the structure of human participation behind pull request activity changes after agent adoption, providing repository-based estimates of how agents affect human participants.
  \item \textbf{Evidence on newcomer relative participation.} We estimate how the share of newcomers among human contributors changes after agent adoption and detect no decrease larger than 2.4 percentage points.
  \item \textbf{Measurement validity of repository metrics under agent adoption.} We show three things: control groups defined by absence from an agent dataset are often treated; participation metrics built from pull requests and accounts may change when agents submit work under developer accounts; and the share of pull requests recorded in GHArchive has declined since mid-2025. We release the detection rules for agent traces.
  \item \textbf{Reproducibility.} The replication package~\cite{zhang2026replication} includes all analysis scripts and data, as well as the analysis decisions recorded before estimation.
\end{itemize}

The remainder of this paper is structured as follows. Section~\ref{sec:Background} provides background on AI coding agents, human participation in open-source projects, and staggered difference-in-differences estimation. Section~\ref{sec:relatedwork} reviews related work. Section~\ref{sec:method} describes the research design, including dataset construction, agent-trace screening, matching, and causal inference strategy. Section~\ref{sec:results} presents the results for each research question. Section~\ref{sec:discussion} discusses implications, methodological lessons, threats to validity, and directions for future work. Section~\ref{sec:conclusion} concludes.

\section{Background}
\label{sec:Background}

\subsection{AI Coding Agents and Their Footprint on GitHub}
\label{sec:2.1}
\paragraph{Definition and scope.}
We adopt the definition of Robbes et al.: an AI coding agent is a large language model running in a loop with tool access that aims to complete a given goal~\cite{robbes2026agentic}. Compared with standalone large language models, agents can perceive and use external resources and tools~\cite{liu2024large}, and can therefore generate complete pull requests starting from a task description provided by a developer~\cite{li2025rise, robbes2026agentic}. Code completion in the IDE and chat assistants fall outside the scope of this paper. Code completion offers suggestions at the cursor in the editor, and the developer decides whether to accept them~\cite{barke2023grounded}; such assistance leaves no explicit traces in the repository~\cite{robbes2026agentic}. With chat assistants, developers copy the generated code into the project and write it to the repository in their own commits; unless the developer states it explicitly, such code cannot be distinguished from code written by humans~\cite{grewal2024analyzing}.

\paragraph{How agents submit their work.}
Common AI coding agents submit their work to GitHub in two ways (Table~\ref{tab:agent-submission}). The first opens pull requests under the agent's own account: for example, the GitHub Copilot cloud agent (formerly coding agent) uses the \texttt{copilot-swe-agent} account~\cite{github2026copilotcloudagentapi}, and Devin uses the \texttt{devin-ai-integration[bot]} account~\cite{li2025rise}. The second submits under the account of the developer who started the task, so the author field is the same as that of a human pull request~\cite{ghaleb2026fingerprinting}. For example, OpenAI Codex and Claude Code attribute their contributions to the developer in commit metadata~\cite{li2025rise}. The submission mode can also change with how an agent is invoked: Cursor's cloud agent opens pull requests under the developer's account when the developer starts it interactively, and under \texttt{cursor[bot]} when it is started by an automation or a service account~\cite{cursor2026cloudagentattribution}. Besides the account, agents leave other markers by default, including branch names that start with the agent's name, such as \texttt{codex/}, \texttt{copilot/}, and \texttt{cursor/}~\cite{li2025rise}, and co-author information in commits or pull requests~\cite{li2025rise, robbes2026agentic}. These markers are default settings that users can turn off; for example, the co-author statement of Claude Code can be disabled~\cite{li2025rise, anthropic2026claudecodesettings}.

\begin{table}[t]
\centering
\caption{Pull request author account and default markers of common AI coding agents.}
\label{tab:agent-submission}
\small
\begin{tabular}{p{0.20\linewidth}p{0.30\linewidth}p{0.30\linewidth}p{0.10\linewidth}}
\toprule
Agent & Pull request author & Default markers & Sources \\
\midrule
GitHub Copilot cloud agent & Agent account (\texttt{copilot-swe-agent}) & \texttt{copilot/} branch prefix; agent authorship in commit metadata & (\cite{github2026copilotcloudagentapi, li2025rise}) \\
Devin & Agent account (\texttt{devin-ai-integration[bot]}) & Agent authorship in commit metadata & (\cite{li2025rise}) \\
OpenAI Codex & Developer account & \texttt{codex/} branch prefix; no attribution in commit metadata & (\cite{li2025rise}) \\
Cursor (cloud agent) & Developer account when started interactively; \texttt{cursor[bot]} when started by an automation or a service account & \texttt{cursor/} branch prefix; agent authorship in commit metadata & (\cite{cursor2026cloudagentattribution, li2025rise}) \\
Claude Code & Developer account & Co-author trailer in commits; generation statement in the pull request description; both can be disabled & (\cite{li2025rise, anthropic2026claudecodesettings}) \\
\bottomrule
\end{tabular}
\end{table}

\paragraph{Observation boundary.}
Repository data therefore allow us to observe agent use that leaves markers, not all use of a given product. When developers use an agent locally and then commit the changes themselves, the traces remain mainly in commits or configuration files and do not necessarily appear in pull requests. A pull-request-based census misses 79\% of the Claude Code adopters detected through commits~\cite{khosravani2026detecting}. Product names cannot define the boundary either: Copilot refers both to inline code completion in the IDE~\cite{barke2023grounded, li2025rise} and to the cloud agent that submits under the \texttt{copilot-swe-agent} account~\cite{github2026copilotcloudagentapi}. In this paper, Copilot refers to the latter (Section~\ref{sec:datacollection}). Sections~\ref{sec:screening} and~\ref{sec:threats} discuss how this boundary affects adoption identification and estimation, respectively.

\subsection{Human Participation in Open-Source Projects}
\label{sec:huamn_in_osp}
\paragraph{Pull-based development.}
On GitHub, open-source projects commonly adopt the pull-based development model: developers modify code in their own copy of the repository and submit the changes to the project as a pull request, and the project's integrators review it and decide whether to merge it~\cite{gousios2014exploratory}. This model has become widely popular in open-source projects~\cite{gousios2016work} and provides more opportunities for community engagement~\cite{gousios2014exploratory}. Each pull request records the account that submitted it, so the pull request history of a project records both who participates and how often.

\paragraph{Unequal participation.}
Participation in open-source projects is highly unequal. The core-periphery model holds that a small group of core developers contributes most of the code and steers the design and evolution of the project, while the periphery consists of co-developers and active users~\cite{crowston2006core}. Empirical studies show that core developers make up only a small fraction of all contributors~\cite{mockus2002two, crowston2006core}, whereas many developers make only occasional contributions~\cite{pinto2016more}. The number of people who participate in a project and the amount they contribute are therefore two distinct dimensions that can change independently of each other.

\paragraph{Newcomers.}
A newcomer is a person who makes a first contribution to a given project within a given period~\cite{chaoss2026newcontributors}. The CHAOSS community health metrics treat metrics related to newcomers as indicators of project health and distinguish the types of their contributions, such as pull requests, commits, issues, and comments~\cite{goggins2021open}.

\paragraph{Accounts and contributors.}
Counting contributors from repository data rests on the assumption that one account corresponds to one person. Prior work identifies two situations in which this assumption does not hold. First, the same developer may use several identities, which need to be merged before counting~\cite{goeminne2013comparison}. Second, activity generated by bot accounts mixes with human activity and needs to be identified separately~\cite{dey2020detecting, golzadeh2021ground}. Agent pull requests submitted under developer accounts, described in Section~\ref{sec:2.1}, do not satisfy this assumption either.

\subsection{Estimating Effects from Staggered Adoption}

In observational data, whether a repository adopts an agent is not randomly assigned, so the effect of adoption cannot be measured directly as the difference between adopting and non-adopting repositories. The idea of difference-in-differences (DiD) is to use the change in the outcome of the control group over the same period to approximate the change that the treatment group would have experienced had it not adopted, that is, the counterfactual; the difference between the actual change of the treatment group and this counterfactual is attributed to the treatment~\cite{roth2023s}. This attribution rests on two assumptions. The first is parallel trends: had the treatment group not adopted, its outcome would have changed along the same trend as that of the control group. Because this assumption concerns the counterfactual outcomes of the treatment group after adoption, it cannot be tested directly. The second is no anticipation: the treatment does not affect the outcome before it is actually implemented. If the recorded adoption time is later than the actual adoption, the data will show the same pattern as anticipation; Section~\ref{sec:causal} describes how this paper addresses this.

In this paper, repositories adopt agents in different months, that is, adoption is staggered. Such data have traditionally been analyzed with two-way fixed effects (TWFE) regressions, which add a treatment indicator to unit fixed effects and time fixed effects. The TWFE estimator can be decomposed into a weighted average of all two-group, two-period DiD comparisons, including comparisons that use already-treated units as controls for later-treated units. If treatment effects change with time since adoption, changes in the effects of the already-treated units are subtracted from the estimate~\cite{goodman2021difference}; some group-period treatment effects therefore receive negative weights, and the regression coefficient may even have the opposite sign of all treatment effects~\cite{de2020two, borusyak2024revisiting}. Dynamic TWFE specifications that include leads and lags of the treatment suffer from the same problem: the coefficient for a given relative period may be contaminated by effects from other periods~\cite{sun2021estimating}.

Recently proposed estimators that are robust to treatment effect heterogeneity share a common feature: they use only comparisons between treated units and not-yet-treated (or never-treated) units, and aggregate these comparisons into a target parameter with explicit weights~\cite{roth2023s}. To describe such estimators, treated units are usually grouped into cohorts $g$ by their adoption time (in this paper, the adoption month), and the relative time $e = t - g$ denotes the number of months since adoption. On this basis, Callaway and Sant'Anna~\cite{callaway2021difference} estimate the average treatment effect of each cohort in each period; Sun and Abraham~\cite{sun2021estimating} estimate the average treatment effect of each cohort at each relative time, which they call the cohort-specific average treatment effect on the treated (CATT); and Borusyak et al.~\cite{borusyak2024revisiting} first impute the counterfactual outcomes from untreated observations and then aggregate. The target parameter obtained by aggregation can be, for example, the average treatment effect over a number of months after adoption (ATT). The estimators differ mainly in which units they use as controls, which pre-treatment periods they use to construct the differences, and which weights they use for aggregation~\cite{roth2023s}. This paper uses the Sun--Abraham estimator; its specification and target parameter are described in Section~\ref{sec:causal}.

Arranging the estimated coefficients for each relative time by $e$ yields an event study. All coefficients are normalized relative to a reference period (usually $e = -1$), so each coefficient indicates how much larger or smaller the gap between the two groups is in that period than in the reference period. The pre-adoption coefficients are commonly used to diagnose parallel trends: if they are close to zero and show no systematic trend, the two groups changed in step before adoption. This diagnostic, however, has two limitations~\cite{roth2022pretest}. First, pre-trend tests may have low power, so failing to reject the null hypothesis does not imply that parallel trends hold. Second, conditioning the analysis on passing the test may exacerbate the bias of point estimates and make the coverage of confidence intervals fall below the nominal level. Rambachan and Roth~\cite{rambachan2023more} therefore propose not to require that parallel trends hold exactly, but to restrict how large the post-adoption violations can be relative to the pre-adoption differences in trends, and to examine whether the conclusions still hold under different restrictions. The sensitivity analysis used in this paper is described in Section~\ref{sec:causal}.

\section{Related Work}
\label{sec:relatedwork}
We review four lines of work related to this study: empirical research on AI coding agents in open-source development (Section~\ref{sec:3.1}), human contributor ecosystems in open-source software (Section~\ref{sec:3.2}), causal inference in software engineering (Section~\ref{sec:3.3}), and the detection of bots and agents in repository data (Section~\ref{sec:3.4}).

\subsection{AI Coding Agents in Open-Source Development}
\label{sec:3.1}

Coding agents have become regular contributors to open-source software. Projects at all stages of maturity use them, including projects of established organizations~\cite{robbes2026agentic}. In snapshots from December 2024 to April 2026, agents made more than 320,000 commits per month~\cite{khosravani2026detecting}. A first group of studies examines the code that agents write~\cite{ogenrwot2026ai} and finds its quality mixed. On the one hand, lines written by agents are changed less often than lines written by humans~\cite{rahman2026will}. On the other hand, agent code needs more bug fixes and has more security weaknesses~\cite{xia2026violent, takerngsaksiri2026finishes}. Someone has to do this extra work, and studies differ on who does it. Human developers make most later changes to files generated by agents~\cite{sawada2026maintenance}, whereas agents write most of the fixes to their own merged pull requests~\cite{takerngsaksiri2026finishes}. What reduces this burden is review: it is lower in projects that review changes before merging them~\cite{xia2026violent}.

A second group of studies therefore examines how people review and accept agent pull requests. People first decide which pull requests to accept. This decision depends strongly on the type of task and on who submits the pull request~\cite{pinna2026comparing}, and people reject pull requests that are large, unwanted, or incorrect~\cite{ehsani2026ai, abujadallah2026understanding}. What they accept, however, they often check little. AI-co-authored pull requests are merged quickly with little feedback~\cite{gao2026autopilot}, and most agent pull requests get no review at all~\cite{duma2026reviews}. When people do engage, the nature of their work changes. They spend less effort writing code and more effort supervising the agent~\cite{khelifi2026behind}, and their reviews take more rounds~\cite{zhong2026synergy}. This supervision starts even before a pull request exists: when users add a requirement after the agent has started editing, about twice as much of the agent's earlier code is deleted or replaced as after other edits~\cite{jiang2026requirements}. These studies therefore show how the work of each person changes. Because they look at single pull requests, however, they cannot show whether the group of people in a project changes.

To see that, studies must compare projects or developers that got an AI tool with similar ones that did not, over time. Studies of GitHub Copilot's code suggestions did this and found that AI changes who does what. Contributions shifted toward maintaining existing code~\cite{yeverechyahu2024impact}, maintainers coded more and managed less~\cite{hoffmann2024generative}, and peripheral developers bore more of the added coordination than core developers~\cite{song2024impact}. Studies of coding agents use the same design but mostly measure output. Development speed rises only briefly or only in some projects, while static analysis warnings and code complexity rise and stay high~\cite{he2026cursor, agarwal2026agents}. The one exception is Xu et al.~\cite{xu2026decoupling}, who study people but only newcomers, and find no drop in their number, the acceptance of their first pull requests, or their retention. Whether agents change the wider group of people behind a project's work therefore remains open. We build on the data and designs of these studies and ask whether the number of distinct people behind each pull request changes after adoption.

\subsection{Human Contributor Ecosystems in Open-Source Software}
\label{sec:3.2}
The continued development of open-source projects depends on a small number of core contributors, and the departure of core contributors is common. Miller et al.~\cite{miller2019people} analyzed why established contributors disengage from open-source projects and found that the most common reasons are transitions such as switching jobs or leaving academia. Avelino et al.~\cite{avelino2019abandonment} found that 16\% of 1,932 popular GitHub projects had been abandoned, and that 41\% of these survived because new core developers took over. Fan et al.~\cite{fan2025core} found that core developer turnover is common in the Rust package ecosystem. These studies focus on the departure of individual contributors within a project, whereas this paper focuses on changes in the participation structure after a project adopts an external tool. Chen et al.~\cite{chen2026does} used a difference-in-differences design with matched control groups to estimate the effect of core contributor disengagement on pull request throughput, acceptance rate, and time to merge. This paper uses a similar quasi-experimental design, but the treatment event is the adoption of AI coding agents, and the outcome variable is the number of distinct humans behind each pull request.

A steady influx of newcomers is what allows core contributors to be replaced. Steinmacher et al.~\cite{steinmacher2015systematic} identified 15 barriers that newcomers face when joining open-source projects and grouped them into five categories: social interaction, newcomers’ previous knowledge, finding a way to start, documentation, and technical hurdles. Using issue tracking data from Mozilla and Gnome, Zhou and Mockus~\cite{zhou2012make} found that the probability of a newcomer becoming a long-term contributor is associated with the newcomer’s willingness and environment. Studies published in 2026 further examine the retention and growth of newcomers. Ouf and Guizani~\cite{ouf2026same} compared the retention of newcomers who joined through events such as Google Summer of Code with that of newcomers who joined on their own. Ouf et al.~\cite{ouf2026good} compared the pathways by which newcomers become core contributors in open-source projects for social good and in conventional open-source projects. Hao et al.~\cite{hao2026first} estimated the effect of the characteristics of a newcomer’s first issue on long-term retention. These studies take the individual newcomer as the unit of analysis and form the starting point of RQ2; this paper instead estimates, at the project level, how the share of newcomers among human contributors changes before and after agent adoption. Hoshikawa et al.~\cite{hoshikawa2026longitudinal} found that, across 37 popular repositories, the proportion of issues labeled good first issue declined significantly from January 2024, and the merge rate of newcomers’ pull requests addressing such issues fell from 61.9\% to 42.2\%. That study describes the aggregated time trend of these projects and does not distinguish whether projects adopted AI tools; this paper identifies the adoption month of each repository and compares it with matched control repositories.

The structure of the contributor population is also related to project sustainability. Gender and tenure diversity are significant positive predictors of the productivity of GitHub teams~\cite{vasilescu2015gender}. The number of a project’s ties and its relative position in the dependency network affect whether it remains active~\cite{valiev2018ecosystem}. Socio-technical network features can predict whether Apache incubator projects become sustainable~\cite{yin2021sustainability}, and participation during a project’s first three months can predict whether it remains active over the following two years~\cite{xiao2023early}. These studies provide the basis for taking the human contributor ecosystem as the dependent variable of this paper.

Afroz et al.~\cite{afroz2026ai} used 2025 as a common intervention point for all repositories and found, across 294 popular repositories, that pull request volume increased while merge rates declined, with the merge rate of one-time contributors 18.18\% below the counterfactual prediction. This paper identifies the agent adoption month of each repository and uses matched controls to separate the effect of adoption from common trends in the same period. Hora et al.~\cite{hora2026we} analyzed 281 AI contribution policies of open-source projects. They found that most policies permit AI-assisted contributions but attach conditions, that common countermeasures include closing pull requests, banning users, and disallowing fully autonomous agents, and that no cross-project convention for disclosing AI use has emerged. That study characterizes the rules projects set for AI contributions, whereas this paper estimates the change in participation structure that actually follows agent adoption. Its finding on disclosure is consistent with the measurement limitation this paper encounters in RQ3.

\subsection{Causal Inference in Software Engineering}
\label{sec:3.3}
Researchers frequently use observational repository data to investigate the effects of development practices and tools, but associations alone do not establish causality. Siebert~\cite{siebert2023causal} conducted a systematic mapping study of software engineering research applying Pearl's graphical framework for statistical causal inference and found that its application remained relatively recent and the research community fragmented. Vegas et al.~\cite{vegas2026causality} examined how retrospective cohort studies, a design established in epidemiology, can be applied in mining software repositories to support causal conclusions from observational comparisons.

Quasi-experimental studies have also examined changes following adoption of automation tools. Wessel et al.~\cite{wessel2022quality} applied regression discontinuity analysis to code review bot adoption in 1,194 GitHub projects. They reported more merged pull requests, fewer non-merged pull requests, and reduced developer communication. Their work demonstrates the value of examining both activity volumes and collaborative practices when studying development automation.

More recent work investigates AI-assisted software development. Agarwal et al.~\cite{agarwal2026ai} used propensity-matched controls and staggered difference-in-differences (DiD), estimated with an imputation-based estimator, to assess the effects of coding agent adoption on repository-level development velocity and code-quality metrics. They found substantial, front-loaded velocity gains only in repositories where the agent was the first observable AI tool; repositories with prior AI IDE usage showed no significant change in velocity. Cognitive complexity increased significantly in both groups, whereas static-analysis warnings increased significantly only for agent-first repositories. Hamza et al.~\cite{hamza2026working} used a developer-level longitudinal comparison, including DiD with control developers from repositories without agent activity, to examine coding, testing, and review behaviors. They associated agent adoption with longer pull request descriptions and increased testing activity, with stable coding effort, whereas controls also showed faster and less thorough reviews, which they attributed to broader ecosystem trends. Xu et al.~\cite{xu2026decoupling} investigated newcomer participation and code complexity using DiD against matched non-adopting controls and found no statistically significant decline in newcomer inflow, despite modest increases in measured complexity. This latter result concerns newcomers rather than all human contributors, and is currently available as a preprint.

Our study complements these findings by focusing on repository-level human participation following agent adoption, including both the scale and relative composition of human contributions. We estimate adoption-cohort-specific effects over time using matched controls and a staggered DiD design.

\subsection{Bot and Agent Detection in Repository Data}
\label{sec:3.4}
Distinguishing human from automated contributions is essential for repository-based studies of developer activity. Dey et al.~\cite{dey2020detecting} proposed BIMAN, which identifies commit bots using author names, commit messages, and commit-related features. Golzadeh et al.~\cite{golzadeh2021groundtruth} developed a classification model based on repetitive patterns in GitHub issue and pull request (PR) comments. More recently, Chidambaram et al.~\cite{chidambaram2025bot} introduced BIMBAS and its implementation, RABBIT, using GitHub activity sequences to identify bots across different activity types.

Detecting AI coding agents presents additional challenges because agent-assisted contributions do not necessarily originate from identifiable bot accounts. Li et al.~\cite{li2026aidev} introduced AIDev, a large-scale dataset of agent-attributed PRs, providing a basis for studying coding agent activity in GitHub repositories. Khosravani and Mockus~\cite{khosravani2026detecting} combined configuration files, commit metadata, author identities, and bot signatures to identify agent traces. Their findings show that PR-based and commit-based detection capture substantially different populations of agent-using repositories.

These studies highlight the distinction between observable agent activity and actual adoption. In particular, the first detected agent PR may occur after a repository has begun using AI coding agents. This limitation matters for our study because inaccurate adoption timing can affect treatment classification and the interpretation of pre-adoption trends in causal analysis.

\section{Research Design}
\label{sec:method}
\begin{figure*}[htbp]
\centering
    \includegraphics[width=1\linewidth]{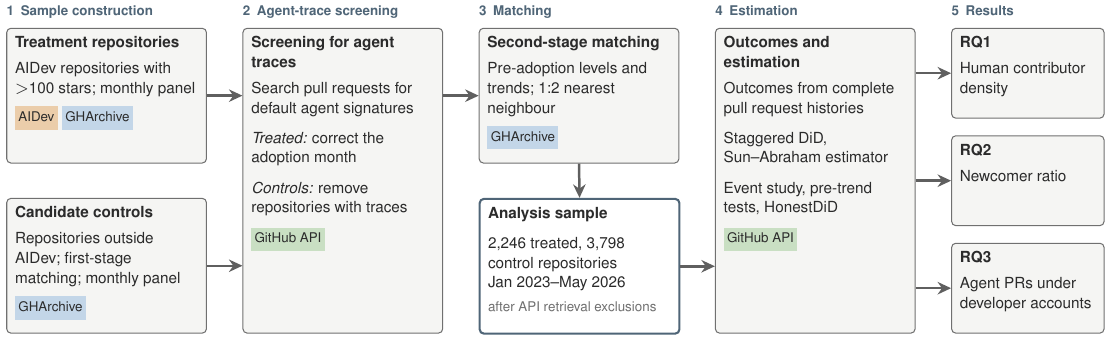}
\caption{Research Design Overview.}
\label{fig:overview}
\end{figure*}

Figure~\ref{fig:overview} gives an overview of our research design, which consists of five stages. First, we take repositories with agent pull requests in AIDev as the treatment group and build a candidate control pool from repositories outside AIDev (Sections~\ref{sec:dataset}–\ref{sec:control}). Second, we scan the pull requests of both groups for the default traces that agents leave, use them to correct the adoption month of treatment repositories, and remove candidate control repositories with traces (Section~\ref{sec:screening}). Third, we perform a second-stage matching on pre-adoption activity levels and trends to obtain the final analysis sample (Section~\ref{sec:second_stage}). Fourth, we define the outcome variables and estimate the effects of adoption using a staggered difference-in-differences design (Sections~\ref{sec:outcome_var}–\ref{sec:causal}). Fifth, the results for each research question are reported in Section~\ref{sec:results}. The three data sources serve different purposes: AIDev is used to identify the treatment group, GHArchive to construct the sample and compute the matching covariates, and the GitHub API to screen for agent traces and compute all outcome variables.

\subsection{Dataset and Treatment Identification}
\label{sec:dataset}
We build our treatment group using AIDev dataset~\cite{li2026aidev}
for three reasons. 
First, it is the first large-scale, publicly available dataset capturing agent-authored pull requests from real-world GitHub repositories, covering five AI coding agents: Claude Code, Cursor, Devin, GitHub Copilot, and OpenAI Codex.
Second, it is the official dataset of the MSR 2026 Mining Challenge, which has been used in multiple peer-reviewed studies and allows our results to be compared with theirs.
Third, it provides large-scale coverage of agent-authored pull requests across a diverse set of repositories, so that adopting repositories can be identified from a shared, documented source.
We include all five agents rather than a single one, because the study concerns the adoption of autonomous AI coding agents as a class of tools, not the effects of a particular product. Additionally, a single agent would yield too few repositories: Devin, for example, has 288 repositories that meet our quality threshold.

We define a repository as a treatment candidate if it satisfies the following quality criteria: (1) it appears in the AIDev dataset with at least one agent-authored pull request, and (2) it has more than 100 GitHub stars at the time of data collection. The stars threshold excludes most personal and toy projects. This threshold is consistent with prior work: the AIDev curated subset itself applies a 100-star threshold~\cite{li2026aidev}, and Agarwal et al. apply a 10-star threshold in a closely related study~\cite{agarwal2026agents}, suggesting that star-based filtering is an established practice in this line of research. Applying these criteria yields 3,124 repository-agent records, corresponding to 2,836 unique repositories. The larger record count reflects the fact that some repositories have adopted multiple agents.

Table~\ref{tab:agent_repos} reports the number of quality-threshold repositories per agent. OpenAI Codex has the largest presence with 1,248 repositories, followed by GitHub Copilot with 1,024, Cursor with 327, Devin with 288, and Claude Code with 237. The most common primary programming languages are TypeScript, Python, and Rust.

\begin{table}[h]
\centering
\caption{Number of repositories meeting quality threshold ($stars > 100$) per AI coding agent}
\label{tab:agent_repos}
\begin{tabular}{lrr}
\toprule
\textbf{Agent} & \textbf{Total Repositories} & \textbf{Repositories ($stars > 100$)} \\
\midrule
Claude Code    & 1,909  & 237   \\
Copilot        & 14,465 & 1,024 \\
OpenAI Codex   & 84,551 & 1,248 \\
Cursor         & 12,669 & 327   \\
Devin          & 4,747  & 288   \\
\midrule
\textbf{Total (unique)} & \textbf{116,211} & \textbf{2,836} \\
\bottomrule
\end{tabular}
\end{table}

We define the adoption month of each repository as the calendar month of its first agent-authored pull request in AIDev, across all agents. Adoption is concentrated in a short period: only 197 repositories adopted agents before May 2025, and most adopted between May and July 2025, with 1,054 in June 2025 alone. 
All 2,836 candidate repositories have their first agent-authored pull request after January 2024, so each has at least twelve months of observation before adoption within our window, which starts in January 2023. Because a repository may have used an agent before its first pull request recorded in AIDev, we also searched the pull requests of each repository for agent signatures (Section~\ref{sec:screening}) and set the adoption month to the earlier of the month of its first agent pull request in AIDev and the month of the first signature found in our scan. This moved the adoption month earlier for 147 of the 2,808 repositories retained in Section~\ref{sec:datacollection} (5.2\%): by one month for 69 repositories, by two to three months for 46, and by more than three months for 32.

\subsection{Data Collection and Panel Construction}
\label{sec:datacollection}
We use GHArchive to construct the sample and to compute the matching covariates (Section~\ref{sec:second_stage}).
GHArchive is a public dataset that records the public event stream of GitHub repositories from 2011 to the present, including events such as pull request creation, merging, and closing.
It is hosted on Google BigQuery and supports large-scale SQL queries. Compared to calling the GitHub REST API repository by repository, using GHArchive enables the extraction of historical data spanning multiple years across thousands of repositories within minutes, while avoiding the constraints of API rate limits.

For the 2,836 candidate repositories, we executed a SQL query in BigQuery to extract all events of type PullRequestEvent with action opened from GHArchive for the period from January 2023 to May 2026, that is, all newly opened pull requests. We aggregated events by repo\_name and calendar month, counting the number of human contributors per repository per month (using distinct actor.login counts) and the total number of pull requests.

For bot filtering, we excluded accounts whose login names contain the strings ``[bot]'' or ``bot'', as well as three commonly used automation accounts: github-actions, dependabot, and renovate. This filtering approach is consistent with the practices adopted in existing related studies~\cite{he2026cursor}.

The query returned 58,848 repo-month records covering 2,808 unique repositories. 
An additional 28 candidate repositories had no human pull request records in GHArchive throughout the observation period. Because they provide no pre-adoption information for the matching covariates, they would also be excluded by the activity requirement in Section~\ref{sec:second_stage}; we therefore excluded them at this stage, and the treatment group comprises 2,808 repositories.
Because repositories without pull request activity in a month have no corresponding records in GHArchive, we expanded the results into a balanced panel covering all 41 months from January 2023 to May 2026 for every repository, filling \texttt{human\_contributors} and \texttt{pr\_count} with zero for months without pull request activity. This panel contains 115,128 repo-month observations (2,808 repositories multiplied by 41 months).

To obtain the newcomer counts used as matching covariates (Section~\ref{sec:second_stage}), we executed a second query against GHArchive in BigQuery. The identification of newcomers depends on the complete contribution history of each contributor within a given repository. To mitigate left-truncation bias, whereby contributors who had already contributed prior to the start of the observation window might be incorrectly classified as newcomers, we set the time range of this query to January 2020 through May 2026, three years earlier than the start of the main analysis window in January 2023. The specific logic is as follows: for each contributor in each repository, we identified the date of their first PR submission to that repository, then aggregated by repository and calendar month to count the number of contributors making their first contribution in each month. The query returned 59,698 repo-month records, saved as the \texttt{newcomer\_monthly} table.

To measure review activity, which we use as a matching covariate (Section~\ref{sec:second_stage}), we executed a third query,
extracting events of type PullRequestReviewEvent and PullRequestReviewCommentEvent from GHArchive for the period January 2023 through May 2026, consistent with the first query. 
We aggregated by repository and calendar month and counted the number of reviews per repository per month, hereafter referred to as review activity.
The query returned 48,418 repo-month records, saved as the \texttt{review\_monthly} table. 

The \texttt{newcomer\_monthly} and \texttt{review\_monthly} tables
were merged into the panel from the first query via left join, using \texttt{repo\_name} and calendar month as the join keys, yielding \texttt{panel\_dataset\_full}.
For repo-month observations with no corresponding records, newcomers and review activity were filled with zero, for the same reasons as the treatment of \texttt{human\_contributors}.
The merged panel retains 115,128 repo-month observations, with two additional columns: the number of newcomers and review activity This GHArchive panel is used only for sample construction and the matching covariates.

We compared the pull request counts in GHArchive with the results of retrieving pull requests repository by repository through the GitHub GraphQL API, and found that GHArchive misses an increasing share of pull requests from mid-2025 onward. Between January 2023 and April 2025, the number of pull requests recorded in GHArchive is about 86\% of the number seen through the API; after that, the share declines month by month, to 32\% in October 2025 and 6.5\% in May 2026. Most repositories adopted agents between May and July 2025, so their post-adoption observation period falls almost entirely within this range. The missing records affect pull request counts more than contributor counts: after May 2025, GHArchive records about 35\% of the pull requests and about 50\% of the contributors seen through the API. Human contributor density computed from GHArchive therefore rises mechanically with the share of missing records. For this reason, we retrieved all pull requests since January 2023 for each repository through the GitHub GraphQL API and used them to compute all outcome variables. Therefore, as we have mentioned above, GHArchive is used only for sample construction and the computation of the matching covariates.

For each pull request, we record its creation time, the author account, and the account type. A pull request is treated as a bot pull request if the author account type is Bot or if the login name matches the rule above. 
This rule also excludes pull requests from the GitHub Copilot coding agent, which the GitHub API reports as a bot account (\texttt{copilot-swe-agent}) but which GHArchive records under the login \texttt{Copilot}, not caught by the name rule alone.
Pull requests whose author account has been deleted account for 0.26\% of all retrieved pull requests; they are excluded from both the pull request count and the contributor count. pr\_count is the number of human pull requests per month, and human\_contributors is the number of distinct accounts that submitted human pull requests in that month, with login names compared case-insensitively. Months without human pull requests are recorded as zero for both.

Newcomers are identified in the same way as in the GHArchive query, with the GitHub API as the data source. To avoid left-truncation bias, we additionally retrieved all pull requests of each repository from January 2020 to December 2022 through the API. For each contributor in each repository, we determine the month of their first human pull request since January 2020; if this month falls between January 2023 and May 2026, the contributor is counted as a newcomer in that month. The bot rule and the handling of login names are the same as in the previous paragraph. Of the contributors active in the final sample from January 2023 onward, 8.8\% had already submitted a pull request to the same repository before 2023 (11.0\% in treatment and 6.9\% in control repositories); without this history, they would have been classified as newcomers.

Of the 6,396 repositories in the main matched sample (Section~\ref{sec:second_stage}), 194 have been deleted or made private and could not be retrieved through the API (27 treatment and 167 control repositories). 
Another 56 repositories (8 treatment and 48 control) have no pull request history on GitHub that corresponds to the GHArchive records: between January 2023 and April 2025, GHArchive recorded at least 10 human pull requests, while the API retrieved fewer than half as many, and two independent retrievals three weeks apart gave the same result. 
We manually inspected two such repositories and confirmed that all of their pull requests had been deleted. 
All of these repositories are excluded from the analysis. As a result, 44 treatment repositories lost all of their matched controls and 35 control groups lost their treatment repository; these are removed as well, and in the 164 groups that lost one control, the remaining control receives the full weight of the group. The final sample contains 2,246 treatment repositories and 3,798 control repositories. 
With the covariates recomputed from the API data (review activity still from GHArchive), the standardized mean difference of every matching covariate in this sample is below 0.08.

\subsection{Candidate Control Pool}
\label{sec:control}
To construct the control group, we first identified a pool of candidate control repositories from GHArchive. We selected repositories that met the following quality criteria during the observation period: at least 20 pull requests (n\_pulls), at least 10 distinct users involved (users\_involved), and at least 100 total events (total\_events). Repositories that appeared in the AIDev dataset were excluded, as these had adopted at least one AI coding agent. From the repositories satisfying these criteria, we randomly sampled 100,000 as the candidate control pool.

We first performed a first-stage screening of the candidate pool to identify repositories with activity levels comparable to the treatment group, and then extracted monthly panel data for these repositories. Specifically, we estimated propensity scores~\cite{rosenbaum1983central} using logistic regression with six aggregate activity covariates (\textit{users\_involved}, \textit{n\_pulls}, \textit{n\_forks}, \textit{n\_comments}, \textit{n\_issues}, and \textit{total\_events}) and, following He et al.~\cite{he2026cursor}, selected up to three nearest neighbors per treatment repository without replacement. This step yielded 8,289 candidate control repositories.
It should be noted that these aggregate covariates were computed over different time windows for the two groups: for treatment repositories, only the pre-adoption period was covered, whereas for candidate repositories, the entire observation period was covered. Moreover, they reflect only the level of activity, not its trend over time. We therefore use this step solely to delimit the candidate set. The final control group is determined after screening the candidates for agent use (Section~\ref{sec:screening}) and a second-stage matching based on the monthly panel (Section~\ref{sec:second_stage}).

Following the same procedure as for the treatment group, we extracted monthly pull request, newcomer, and review data for the 8,289 candidate control repositories, again extending the newcomer query back to January 2020, and expanded the data into a balanced panel of 339,849 repo-month observations (8,289 repositories over 41 months).

\subsection{Screening for Agent Traces}
\label{sec:screening}
Exclusion from AIDev does not guarantee that a candidate repository never used a coding agent: AIDev covers a limited collection period and five agents, and developers may submit agent-generated pull requests from their own accounts. We therefore searched the pull requests of all 8,289 candidate control repositories and all 2,808 treatment repositories for signatures that coding agents leave by default, using the GitHub GraphQL API. We used three kinds of signatures: branch name prefixes (e.g., \texttt{codex/}, \texttt{copilot/}), agent author accounts (e.g., \texttt{copilot-swe-agent}), and footers or links in the pull request description (e.g., ``Generated with Claude Code''). 
We defined 15 signals, of which 14 produced at least one hit; all are listed in the replication package~\cite{zhang2026replication}.
The scan covered March 2024 to May 2026 for candidate control repositories, and March 2024 to the end of the adoption month for treatment repositories. The detection rules and the rules for using the scan results were fixed before we inspected when the traces first appeared and before re-estimating any effect; both are documented in the replication package~\cite{zhang2026replication}.

To assess the precision of each signal, two researchers jointly inspected a random sample of 20 hits per signal (280 pull requests in total), examining the pull request description, commits, and author account. A hit was counted as correct if an AI coding agent created the pull request or contributed changes to it. All 280 sampled pull requests met this criterion; with 20 hits per signal, the lower bound of the 95\% confidence interval for the precision of each signal is 83.2\%. Because the scan detects only agent use that leaves a signature, the shares reported below are lower bounds.

Of the 8,032 candidate control repositories that could still be retrieved, 2,554 (31.8\%) showed at least one agent trace. Table~\ref{tab:contamination} reports when these traces first appeared. By the end of June 2025, in the middle of the period in which most treatment repositories adopted agents, 2.0\% of candidate control repositories showed a trace; the share then rose by 6 to 10 percentage points per quarter and reached 31.8\% by May 2026. We removed all 2,554 repositories from the control pool. The remaining 257 candidates could not be retrieved because they had been deleted or made private; we kept them in the pool and examined the effect of removing them in a robustness check. Of the 2,776 treatment repositories that could be retrieved, 2,750 showed a trace by the end of their adoption month. For 2,603 of them (94.7\%), the first trace fell in the AIDev adoption month; for the remaining 147, it fell earlier, and we corrected their adoption month accordingly (Section~\ref{sec:dataset}).

\begin{table}[tb]
\caption{Cumulative share of candidate control repositories with a first agent trace, by calendar quarter ($n = 8{,}032$ candidate control repositories retrievable via the GitHub API). Q2 2026 covers April and May only.}
\label{tab:contamination}
\centering
\begin{tabular}{lrrr}
\toprule
Quarter & New & Cumulative & Cumulative (\%) \\
\midrule
2024 Q2 & 4   & 4     & 0.0  \\
2024 Q3 & 2   & 6     & 0.1  \\
2024 Q4 & 12  & 18    & 0.2  \\
2025 Q1 & 24  & 42    & 0.5  \\
2025 Q2 & 116 & 158   & 2.0  \\
2025 Q3 & 498 & 656   & 8.2  \\
2025 Q4 & 652 & 1,308 & 16.3 \\
2026 Q1 & 806 & 2,114 & 26.3 \\
2026 Q2 & 440 & 2,554 & 31.8 \\
\bottomrule
\end{tabular}
\end{table}

For RQ3, we additionally scanned the 2,325 treatment repositories of the main matched sample (before the exclusions in Section~\ref{sec:datacollection}) from their adoption month to May 2026, using the same detection rules, 
and recorded for each month the number of pull requests and the number of those opened by bot accounts. Of these repositories, 2,302 could be retrieved; the remaining 23 had been deleted or made private. This additional scan is used only for RQ3 and does not change the screening results or the adoption months reported above.

\subsection{Second-Stage Matching}
\label{sec:second_stage}
After screening, 5,735 candidate control repositories remained in the pool. Because the first-stage covariates capture only activity levels and are measured over different windows for the two groups, we re-matched the treatment repositories using covariates computed from the monthly panel, with identical definitions for both groups. All covariates are computed over the pre-adoption window, from January 2023 to April 2025; for treatment repositories that adopted agents before May 2025, the window ends in the month before adoption, so that its length ranges from 15 to 28 months (median 28 months). 
The covariates fall into two categories. The first consists of five level covariates: the monthly means (log-transformed) of the number of human contributors, pull requests, newcomers, and review activity, and the proportion of months with at least one human pull request. The second consists of three trend covariates: for the monthly values (log-transformed) of the number of human contributors, pull requests, and newcomers, we fit a straight line over time and take its slope. A positive slope indicates that the repository was growing before adoption, and a negative slope indicates that it was declining.

Before matching, we required repositories in both groups to have human pull request activity in at least one month of the pre-adoption window, because repositories without such activity cannot provide pre-adoption information for our ratio-based outcome variables. This criterion excluded 363 treatment repositories, leaving 2,445 eligible treatment repositories and 5,547 eligible control repositories. We then performed 1:2 nearest neighbor matching using the MatchIt package~\cite{ho2011matchit}, that is, for each treatment repository we selected the two control repositories with the closest propensity scores. Matching was performed without replacement, that is, each control repository could be selected at most once. We also applied a caliper, that is, the maximum allowed difference in scores between matched repositories: if the difference between a treatment repository and its closest control repository exceeded this limit, no match was made. Following the recommendation of Austin~\cite{austin2011optimal}, the caliper was set to 0.2 times the standard deviation of the logit of the propensity score. A total of 120 treatment repositories had no control within the caliper and were therefore excluded. The final matched sample comprises 2,325 treatment repositories and 4,071 control repositories. Because some treatment repositories were matched to only one control, all estimations use the matching weights.

\begin{table}[tb]
\caption{Covariate balance before and after propensity score matching (main specification), before the exclusions in Section~\ref{sec:datacollection}}
\label{tab:balance}
\centering
\begin{tabular}{llrr}
\toprule
Covariate & Type & SMD before & SMD after \\
\midrule
Human contributors     & Level & 0.282 & 0.054 \\
Pull requests          & Level & 0.297 & 0.062 \\
Newcomers              & Level & 0.309 & 0.051 \\
Review activity        & Level & 0.288 & 0.049 \\
Share of active months & Level & 0.136 & 0.027 \\
Human contributors     & Trend & 0.437 & 0.001 \\
Pull requests          & Trend & 0.437 & 0.001 \\
Newcomers              & Trend & 0.428 & 0.025 \\
\bottomrule
\end{tabular}
\end{table}

Table~\ref{tab:balance} reports the standardized mean difference (SMD) of each covariate before and after matching. The SMD is the difference between the group means divided by the pooled standard deviation of the two groups, computed on the eligible sample before matching~\cite{austin2011introduction}. Before matching, the treatment repositories and the remaining candidates differed clearly in both levels (SMD between 0.136 and 0.309) and trends (between 0.428 and 0.437). These differences are larger than in the candidate pool before screening, where the level covariates had SMDs of at most 0.100 and the trend covariates between 0.318 and 0.336. The repositories removed by the screening, which later adopted agents, thus resembled the treatment repositories more closely than the repositories without any trace did. After matching, the SMD of every covariate is below 0.07, and below 0.03 for the three trend covariates. Because the remaining candidates differ more from the treatment group, we also report estimates from a design that does not rely on controls that never showed a trace (Section~~\ref{sec:causal}). Balance on these observed covariates does not by itself guarantee that the parallel trends assumption holds; we therefore examine pre-adoption dynamics directly using event study estimates in Section~\ref{sec:results}.

To examine the sensitivity of the results to these design choices, we additionally adopted five alternative matching specifications: raising the activity requirement to at least three and at least six active months, using 1:1 matching, removing the activity requirement, and removing the 257 candidates that could not be retrieved via the API. Under all specifications, the SMD of every covariate is below 0.09, except for the newcomer level covariate under the six-month requirement (0.104).

The final dataset combines the treatment and the control panels built from the GitHub API (Section~\ref{sec:datacollection}), restricted to the matched sample of the main specification. After the exclusions described in Section~\ref{sec:datacollection}, it comprises 6,044 repositories (2,246 treatment and 3,798 control) and 247,804 repo-month observations.

\subsection{Outcome Variables}
\label{sec:outcome_var}
We define two complementary outcome variables for RQ1. The first measure is the log-transformed monthly human contributor count, defined as $\log(\mathit{human\_contributors} + 1)$, where $\mathit{human\_contributors}$ is the number of distinct human accounts that submitted at least one pull request to a given repository in a given month; the additive smoothing term handles zero-activity months. This measure captures changes in the absolute number of human contributors. The second measure is human contributor density ($\mathit{hc\_density}$), defined as $\mathit{human\_contributors} / \mathit{pr\_count}$, that is, the ratio of monthly human contributors to the number of pull requests submitted from non-bot accounts, computed only for months in which $\mathit{pr\_count} > 0$. 
This measure captures the number of distinct human accounts behind each pull request;
it decreases when the number of pull requests grows faster than the number of people
submitting them, for example when agents take over part of the pull requests or when
developers submit more pull requests with the help of agents (Section~\ref{sec:threats}).
We designate human contributor density as the primary outcome and the log-transformed number of human contributors as the secondary outcome.

For RQ2, we also define two complementary outcome variables. The first measure is the log-transformed monthly newcomer count, defined as $\log(\mathit{newcomers} + 1)$, where a newcomer is a user who submits a pull request to a given repository in a given month and has no prior pull request activity in that repository between January 2020 and the month in question. The history from January 2020 avoids classifying contributors who were already active before 2023 as newcomers (Section~\ref{sec:datacollection}). The second measure is newcomer ratio ($newcomer\_ratio$), defined as $newcomers/human\_contributors$, computed only for repository-months in which $human\_contributors>0$. 
This measure captures the share of newcomers among the human contributors active in a month.
Analogously, the newcomer ratio is the primary outcome and the log-transformed number of newcomers the secondary outcome.

RQ3 uses human contributor density as the outcome and groups treatment repositories by how much agent work is recorded as human activity after adoption. For each treatment repository, we use the additional scan (Section~\ref{sec:screening}) to compute the number of pull requests from adoption month to May 2026 that carry an agent signature and were opened by a user account, divided by the number of pull requests opened by non-bot accounts in the same months. The higher this share, the more agent work our panel counts as human pull requests. Because the scan detects only agent pull requests that leave a signature, the share is a lower bound. 
Repositories that used the GitHub Copilot coding agent are excluded from RQ3, as recorded in our analysis plan before the analysis~\cite{zhang2026replication}. The plan was written when our outcome data came from GHArchive, where Copilot pull requests appear under human accounts, whereas the scan counts them on the bot side, so the two counts would not have been consistent. With the API data, Copilot pull requests are removed as bot pull requests from both (Section~\ref{sec:datacollection}); we nevertheless keep the exclusion as recorded.
After this exclusion, 1,019 treatment repositories remain; 8 of them had no pull requests from non-bot accounts after adoption, leaving 1,011 repositories for the analysis.

\subsection{Causal Inference Strategy}
\label{sec:causal}
We employ a staggered difference-in-differences (DiD) design to estimate the causal effect of AI agent adoption on human contributor participation. Because different repositories adopted AI agents in different months, the data exhibit staggered treatment timing. Under treatment effect heterogeneity across cohorts, the traditional two-way fixed effects (TWFE) estimator can produce biased estimates, as early-adopting repositories are incorrectly used as controls for late-adopting ones~\cite{wooldridge2025two}. We therefore adopt the interaction-weighted estimator proposed by Sun and Abraham~\cite{sun2021estimating}, which decomposes the aggregate average treatment effect on the treated (ATT) into a weighted average of cohort-specific average treatment effects (CATTs) at each relative time period.

The model includes repository fixed effects and month fixed effects, absorbing time-invariant repository-level heterogeneity and common time trends shared across all repositories, respectively. 
All estimations are weighted by the matching weights (Section~\ref{sec:second_stage}). Standard errors are clustered at the repository level to account for serial correlation in observations from the same repository over time. The parallel trends assumption cannot be tested directly, because it concerns the counterfactual outcomes of treatment repositories after adoption. We therefore examine pre-adoption dynamics in two ways. First, event study plots display the pre-adoption coefficients and their 95\% confidence intervals. 
Second, a joint Wald test examines whether the coefficients from month $-12$ to the month before the reference month are all zero.
A non-significant result indicates that we find no evidence of a violation of the assumption. 

We define the ATT as the equal-weight average of the event study coefficients for months 1 to 12 after adoption, with the standard error obtained from the covariance matrix of these coefficients. The adoption month is excluded for two reasons. First, treatment repositories are identified by an agent pull request, so they necessarily have pull request activity in the adoption month, whereas control repositories are not conditioned in this way. Second, outcomes are aggregated by calendar month, so the adoption month mixes days before and after adoption. Months after month 12 are excluded because they are observed only for repositories that adopted early. We also report the estimate that includes the adoption month (months 0 to 12); if this estimate is significant while the estimate for months 1 to 12 is not significant or has less than half its absolute value, we consider the former to depend on the adoption month. 

Regarding the reference month, because our adoption date relies on agent use that leaves a record, in AIDev or as a trace in our scan (Section~\ref{sec:dataset}), actual adoption may have started earlier (Section 6.3). This measurement error means that month $-1$ relative to the recorded adoption month may itself be a post-adoption month for some repositories, in which case a decline in month $-1$ reflects anticipation rather than a violation of parallel trends. We therefore adopt the following protocol: we report results under both month $-1$ and month $-2$ as the reference month for each outcome; if the joint Wald test for pre-adoption coefficients fails under month $-1$ but not under month $-2$, and the coefficient for month $-1$ is itself significantly negative, we treat month $-2$ as the primary reference month and interpret this pattern as consistent with early adoption. 
With month $-2$ as the reference month, the coefficient for month $-1$ is estimated and reported. 

We also conduct a sensitivity analysis using the HonestDiD approach~\cite{rambachan2023more} under the relative magnitudes restriction. 
HonestDiD allows the gap between the two groups to change on its own after adoption and measures the magnitude of this change relative to the pre-adoption period. For example, 5\% means that, after adoption, the gap between the two groups changes each month by at most 5\% of the largest month-to-month change before adoption. The target parameter of HonestDiD is the ATT defined above (months 1 to 12 after adoption). 
The pre-adoption months that define this benchmark run from month $-12$ to the month before the reference month; the months between the reference month and month 1 (month 0, and also month $-1$ when month $-2$ is the reference month) are treated as post-adoption months that receive no weight in the target parameter, so the decrease in month $-1$ does not enter the benchmark.
For each magnitude, HonestDiD yields a 95\% confidence interval for this target parameter; as long as the interval excludes zero, the effect remains significantly different from zero at that magnitude. Among the magnitudes tested (0, 5\%, 10\%, \ldots), the largest one at which the effect remains significant is the degree of violation of parallel trends that the result tolerates.

Because the candidates remaining after screening differ more from the treatment group (Section~\ref{sec:second_stage}), we also use a design that does not rely on them: we keep the matched sample from before screening and, for control repositories with an agent trace, use only the months before their first trace. Each control repository is thus used as a control only while no agent use is detected, and the matched sample remains unchanged, except for 38 treatment and 193 control repositories without data from the GitHub API. The design thus covers 2,413 treatment and 4,564 control repositories, of which 1,531 control repositories are used only before their first trace.

Both ratio outcomes (Section~\ref{sec:outcome_var}) are defined only for months with human pull request activity, so months without such activity drop out of their estimation, possibly at different rates in the two groups. We therefore estimate the same model for an indicator of any human pull request activity in a repository-month. As a sensitivity check, we re-estimate the ratio outcomes on the repositories with human pull request activity in at least half of their months 1 to 12 after adoption, keeping only matched sets that still contain the treatment repository and at least one control and recomputing the matching weights; for control repositories, months after adoption are counted from the adoption month of the treatment repository they were matched to. Because this sample is selected on post-adoption activity, it serves only as a sensitivity check, not as an alternative estimate. We consider a ratio outcome robust to months without activity if the estimate on this sample has the same sign as the primary ATT and lies within its 95\% confidence interval. 

For RQ3, we divide the 1,011 treatment repositories into three groups (low, middle, high) at the tertiles of the share defined in Section~\ref{sec:outcome_var} (7.0\% and 33.3\%), with 337, 340, and 334 repositories, and each control repository follows the treatment repository it was matched to. One repository in the high group contributes no observations to the estimation, so Table~\ref{tab:rq3} reports 333 for this group. 
In each group, we estimate the same model as for RQ1 and test the difference between the ATTs of the high and the low group. Because matching was performed without replacement, the two groups share no repositories, and the standard error of the difference is the square root of the sum of the squared standard errors of the two groups. 
We first grouped repositories by the account type of their first agent pull request; because that grouping reflects only the agent used at adoption and its result was inconclusive, we replaced it with the post-adoption share after the additional scan and before estimating the results reported here. Both versions are recorded in the replication package~\cite{zhang2026replication}.
Because the share is measured after adoption, the RQ3 results indicate an association and do not support a causal interpretation.

\section{Results}
\label{sec:results}
\subsection{RQ1: Human Contributor Participation After AI Agent Adoption}
\label{sec:humanpartici}
Table~\ref{tab:desc_rq1} reports the descriptive statistics for the two RQ1 outcome variables in the matched sample, weighted by the matching weights. To compare the two groups over the same time frame, each control repository inherits the adoption month of the treatment repository to which it was matched, and this month is used to divide the pre-adoption and post-adoption periods. That is, if a project started using AI agents in June 2025, the projects matched to it that did not use agents are also divided at June 2025, and their changes before and after that month are compared.
Before adoption, the treatment group had on average 5.239 human contributors per repository per month, compared with 4.073 for the control group. This difference does not contradict the matching results: the matching balances the log-transformed monthly means (SMD of 0.061 in the final sample, Section~\ref{sec:datacollection}), whereas raw means are sensitive to extreme values, and the distributions of both groups are highly right-skewed (standard deviations of 14.333 and 21.300 before adoption, respectively). That is, most projects have only a few people submitting pull requests each month, while a small number of projects have many more, which pulls the averages up; for a typical project, the two groups are similar in the scale of participation. After adoption, the number of human contributors increased to 6.901 in the treatment group and decreased to 3.439 in the control group. In other words, more people submitted pull requests each month in projects that adopted agents and fewer in projects that did not; however, as shown below, the two groups did not follow the same trend before adoption, so this difference cannot be attributed to AI agents. Human contributor density decreased slightly in the treatment group, from 0.478 to 0.463, and increased slightly in the control group, from 0.542 to 0.564. In other words, the number of distinct people behind every 100 pull requests fell slightly in projects that adopted agents and rose slightly in projects that did not.
These raw changes are for descriptive reference only; the DiD estimates below additionally control for repository fixed effects and common time trends, i.e., they remove both the inherent differences between projects and the changes that all projects experienced during the same period.

\begin{table}[t]
\centering
\caption{Descriptive statistics for RQ1 outcome variables (matched sample, weighted)}
\label{tab:desc_rq1}
\begin{tabular}{lllrrr}
\toprule
Outcome & Group & Period & Mean & SD & $N$ \\
\midrule
\multirow{4}{*}{Human contributors}
 & Treated & Pre-adoption  & 5.239 & 14.333 & 64,696 \\
 & Treated & Post-adoption & 6.901 & 17.632 & 27,390 \\
 & Control & Pre-adoption  & 4.073 & 21.300 & 109,600 \\
 & Control & Post-adoption & 3.439 & 14.431 & 46,118 \\
\midrule
\multirow{4}{*}{HC density}
 & Treated & Pre-adoption  & 0.478 & 0.327 & 38,047 \\
 & Treated & Post-adoption & 0.463 & 0.330 & 19,244 \\
 & Control & Pre-adoption  & 0.542 & 0.341 & 61,538 \\
 & Control & Post-adoption & 0.564 & 0.342 & 23,696 \\
\bottomrule
\end{tabular}
\end{table}

\begin{figure}[t]
  \centering
  \begin{minipage}[t]{0.48\linewidth}
    \centering
    \includegraphics[width=\linewidth]{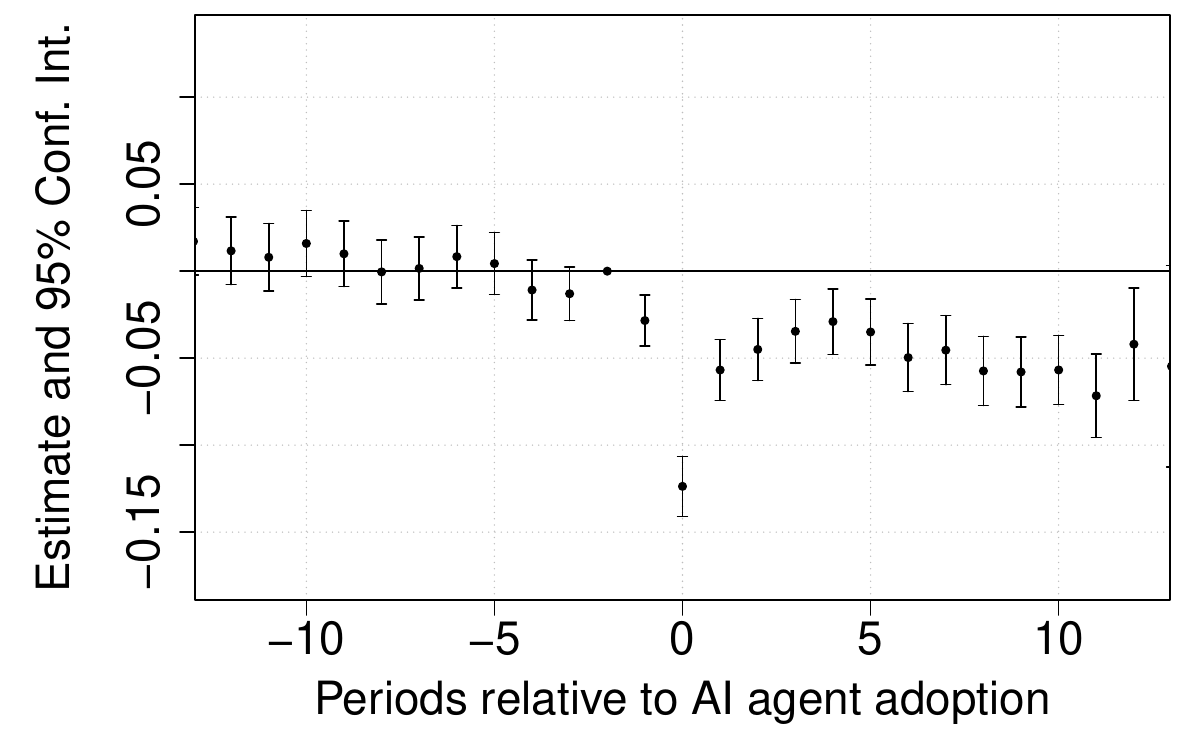}
    \caption{Event study estimates for human contributor density (primary outcome of RQ1). Reference period: month $-2$. Error bars denote 95\% confidence intervals.}
    \label{fig:rq1_density}
  \end{minipage}
  \hfill
  \begin{minipage}[t]{0.48\linewidth}
    \centering
    \includegraphics[width=\linewidth]{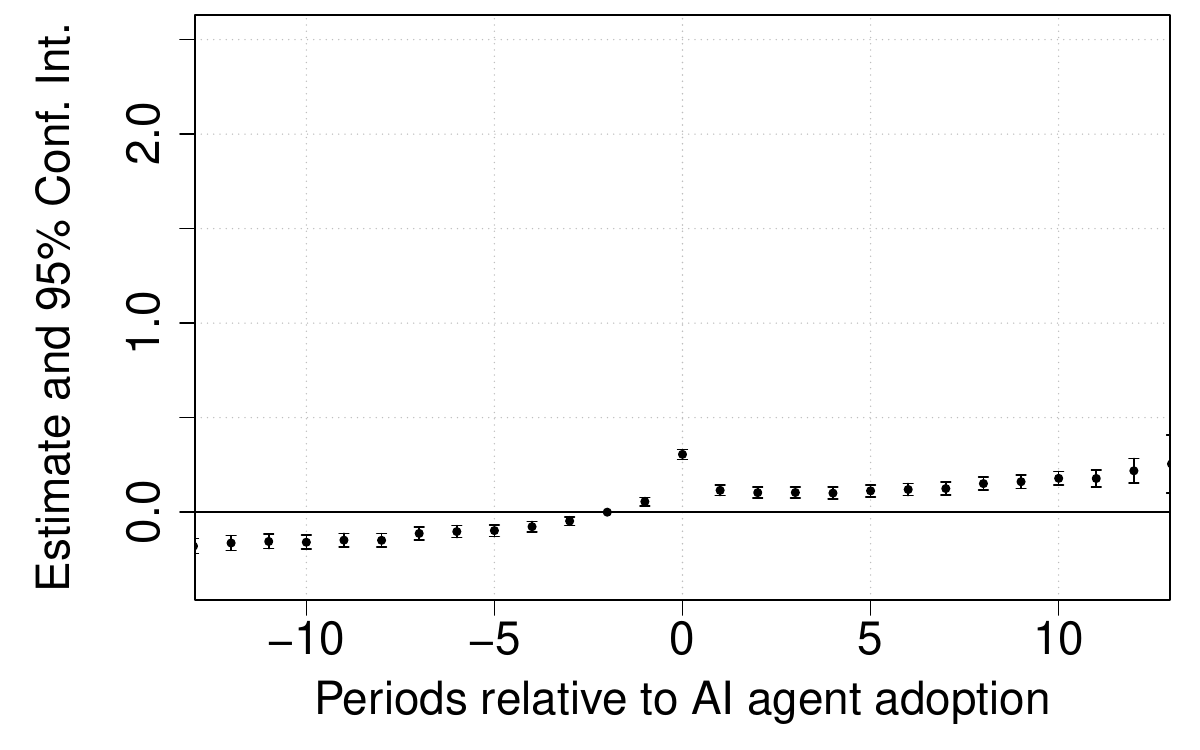}
    \caption{Event study estimates for the log-transformed number of human contributors (secondary outcome of RQ1). Reference period: month $-2$. Error bars denote 95\% confidence intervals.}
    \label{fig:rq1_log}
  \end{minipage}
\end{figure}

In Figures~\ref{fig:rq1_density} and~\ref{fig:rq1_log}, the horizontal axis represents the number of months relative to each repository's AI agent adoption month, where $0$ denotes the adoption month, negative values denote pre-adoption months, and positive values denote post-adoption months; the vertical axis reports the coefficient of the Sun and Abraham estimator for each relative month, representing the average change in the outcome variable in that month relative to the reference month, with error bars denoting 95\% confidence intervals. In other words, each point in the figures shows how much the gap between adopting projects and control projects in a given month before or after adoption is larger or smaller than in the reference month; an error bar crossing zero indicates no confirmable difference between the two groups in that month.

We first use the month before adoption (month $-1$) as the reference month and conduct a joint Wald test of whether the coefficients for months $-12$ to $-2$ are all zero. This test addresses the question of whether the two groups of projects changed in step before adopting agents; only if the two groups were originally in step can differences emerging after adoption be attributed to agents. For human contributor density, the test rejects this null hypothesis ($\chi^2 = 34.10$, $df = 11$, $p < 0.001$). 
However, these pre-adoption coefficients remain stable overall, between $0.015$ and $0.044$, without a systematic upward or downward trend; the rejection arises mainly because month $-1$ itself is lower than the earlier months. 
That is, for most of the pre-adoption period, there is no clear difference in trends between the two groups, except that adopting projects already show a change in the month before adoption. 
This pattern is consistent with anticipation, i.e., some projects may have started using agents before their first recorded agent use.
Following the protocol in Section~\ref{sec:causal}, we therefore use month $-2$ as the reference month. With this specification, density had already declined significantly in month $-1$ ($-0.028$, $p < 0.001$), whereas the joint test of the coefficients for months $-12$ to $-3$ is not significant at the 5\% level ($\chi^2 = 14.88$, $df = 10$, $p = 0.137$), i.e., we find no clear evidence of a violation of parallel trends (Figure~\ref{fig:rq1_density}). 
The coefficients for months $-4$ and $-3$ are slightly negative ($-0.011$ and $-0.013$) but not significant.
In other words, with the second month before adoption as the reference, we find no evidence that the two groups changed out of step before adoption. 

For the log-transformed number of human contributors, the joint test clearly rejects the null hypothesis regardless of whether month $-1$ or month $-2$ is used as the reference month ($\chi^2 = 148.45$ and $85.32$, both $p < 0.001$), and Figure~\ref{fig:rq1_log} also shows systematic deviations in the pre-adoption coefficients. That is, projects that later adopted agents were already changing out of step with control projects in their number of participants before adoption. Therefore, the parallel trends assumption does not hold for this outcome variable, and its estimates cannot be interpreted causally; they are reported below for descriptive purposes only. 

The post-adoption coefficients in Figure~\ref{fig:rq1_density} show the effect of agent adoption on human contributor density.
With month $-2$ as the reference month, the average treatment effect on the treated (ATT), averaged over months 1 to 12 after adoption (Section~\ref{sec:causal}), is $-0.048$ ($p < 0.001$), about 10\% of the pre-adoption level, i.e., the number of distinct people behind every 100 pull requests decreases by about five; with month $-1$ as the reference month, it is $-0.019$ ($p = 0.009$).
Both remain consistent in direction and similar in magnitude across the robustness matching specifications (from $-0.044$ to $-0.059$ and from $-0.015$ to $-0.033$, respectively); with month $-1$ as the reference month, the estimate under 1:1 matching is not significant ($p = 0.057$).
Including the adoption month (months 0 to 12), the estimates are $-0.054$ and $-0.025$ (both $p < 0.001$); excluding the adoption month thus reduces the estimates only slightly, and the result is not driven by the adoption month.
Month by month, the decline is most pronounced in the adoption month ($-0.124$), which is excluded from the ATT for the reasons given in Section~\ref{sec:causal}; it then narrows to about $-0.03$ in months 3 to 5 and widens again to between $-0.042$ and $-0.072$ in months 6 to 12, without returning to the pre-adoption level.
This decline may be partly mechanical, because agent pull requests submitted from developers' own accounts are counted in the denominator (Section~\ref{sec:threats}). 
The HonestDiD sensitivity analysis (Section~\ref{sec:causal}) shows that, with month -2 as the reference month, the effect remains significantly different from zero as long as the post-adoption deviation between the two groups does not exceed 15\% of the largest pre-adoption month-to-month deviation; with month -1 as the reference month, it does not tolerate even a deviation of 5\%.
In a design that keeps the original matched sample and drops control repositories from the month of their first agent trace onward (Section~\ref{sec:causal}), the estimates are similar: $-0.044$ ($p < 0.001$) with month $-2$ and $-0.017$ ($p = 0.011$) with month $-1$ as the reference month; with month $-1$, however, the pre-adoption trend test rejects parallel trends in this design ($p = 0.002$).
Because months without human pull request activity drop out of this ratio (Section~\ref{sec:newcomer}), we also re-estimate it on the repositories that were active in at least half of their months 1 to 12 after adoption (Section~\ref{sec:causal}; 1,176 treatment and 1,473 control repositories): the estimate is $-0.045$ ($p < 0.001$) with month $-2$ as the reference month, within the 95\% confidence interval of the primary ATT, and $-0.012$ ($p = 0.15$) with month $-1$.

\begin{tcolorbox}[colback=gray!10, colframe=gray!50, boxrule=0.5pt, left=6pt, right=6pt, top=4pt, bottom=4pt]
\textbf{Answer to RQ1:} After AI agent adoption, the relative participation of humans in project contributions shows a sustained decline: the number of distinct people behind each pull request decreases by about 4\% to 10\%, depending on the reference month. 
That is, each human contributor corresponds to more pull requests, which does not necessarily mean that humans do less; whether this reflects agent pull requests recorded under developer accounts could not be determined (Section~\ref{sec:rq3}). The estimate should be read as suggestive evidence.
\end{tcolorbox}

\subsection{RQ2: Newcomer Relative Participation After AI Agent Adoption}
\label{sec:newcomer}

Table~\ref{tab:rq2_descriptive} shows that after adoption the newcomer ratio barely changed in the treatment group (from 0.333 to 0.332, $-0.001$) but declined in the control group (from 0.332 to 0.313, $-0.019$). Taken at face value, these numbers suggest that projects that adopted agents retained their share of newcomers better than comparable projects.
However, the composition of the two groups changed differently after adoption. 
The newcomer ratio can only be computed for months in which humans submitted pull requests, and 17\% of control projects (637 of 3,798) had no human pull request activity after adoption, compared with only 1\% of treatment projects. The event study of an indicator of any human pull request activity in a month (Section~\ref{sec:causal}) shows the same pattern: in months 1 to 12 after adoption, adopting projects are 4.0 to 6.8 percentage points more likely than control projects to have human pull request activity (both $p < 0.001$). Because the pre-adoption trend test fails with either reference month (both $p < 0.001$), this difference cannot be interpreted causally. It nevertheless means that the estimates for both ratio outcomes, in RQ1 and RQ2, apply to months in which projects are active. 
The estimates below compare the change within the same project before and after adoption, and we check their sensitivity to months without activity.

\begin{table}[tb]
\centering
\caption{Descriptive statistics for RQ2 outcome variables (matched sample, weighted)}
\label{tab:rq2_descriptive}
\begin{tabular}{lllrrr}
\toprule
Outcome & Group & Period & Mean & SD & N \\
\midrule
\multirow{4}{*}{Newcomers}
 & Treated & Pre-adoption  & 1.575 & 5.827  & 64,696 \\
 & Treated & Post-adoption & 2.416 & 7.883  & 27,390 \\
 & Control & Pre-adoption  & 2.071 & 20.075 & 109,600 \\
 & Control & Post-adoption & 1.544 & 13.261 & 46,118 \\
\midrule
\multirow{4}{*}{Newcomer ratio}
 & Treated & Pre-adoption  & 0.333 & 0.340 & 38,047 \\
 & Treated & Post-adoption & 0.332 & 0.334 & 19,244 \\
 & Control & Pre-adoption  & 0.332 & 0.381 & 61,538 \\
 & Control & Post-adoption & 0.313 & 0.374 & 23,696 \\
\bottomrule
\end{tabular}
\end{table}

\begin{figure}[tb]
\centering
\begin{minipage}[t]{0.48\textwidth}
  \centering
  \includegraphics[width=\linewidth]{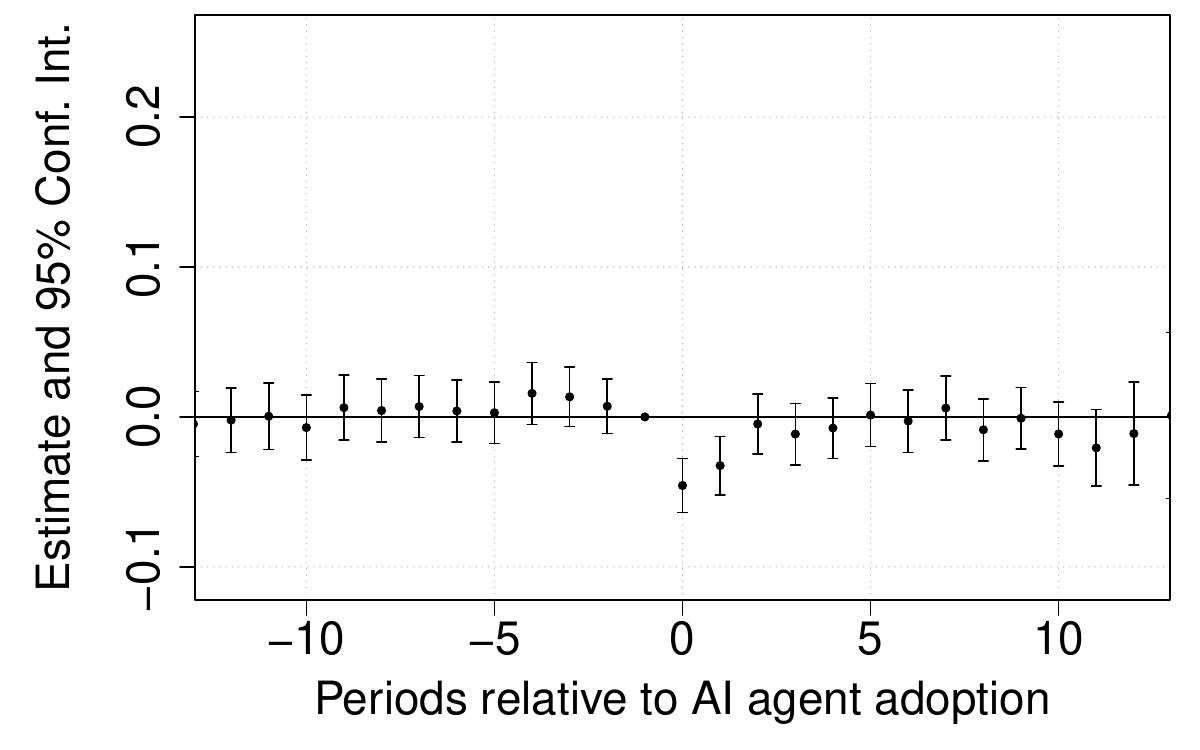}
  \caption{Event study estimates for newcomer ratio (primary outcome of RQ2). Reference period: month $-1$. Error bars denote 95\% confidence intervals.}
  \label{fig:rq2_ratio}
\end{minipage}
\hfill
\begin{minipage}[t]{0.48\textwidth}
  \centering
  \includegraphics[width=\linewidth]{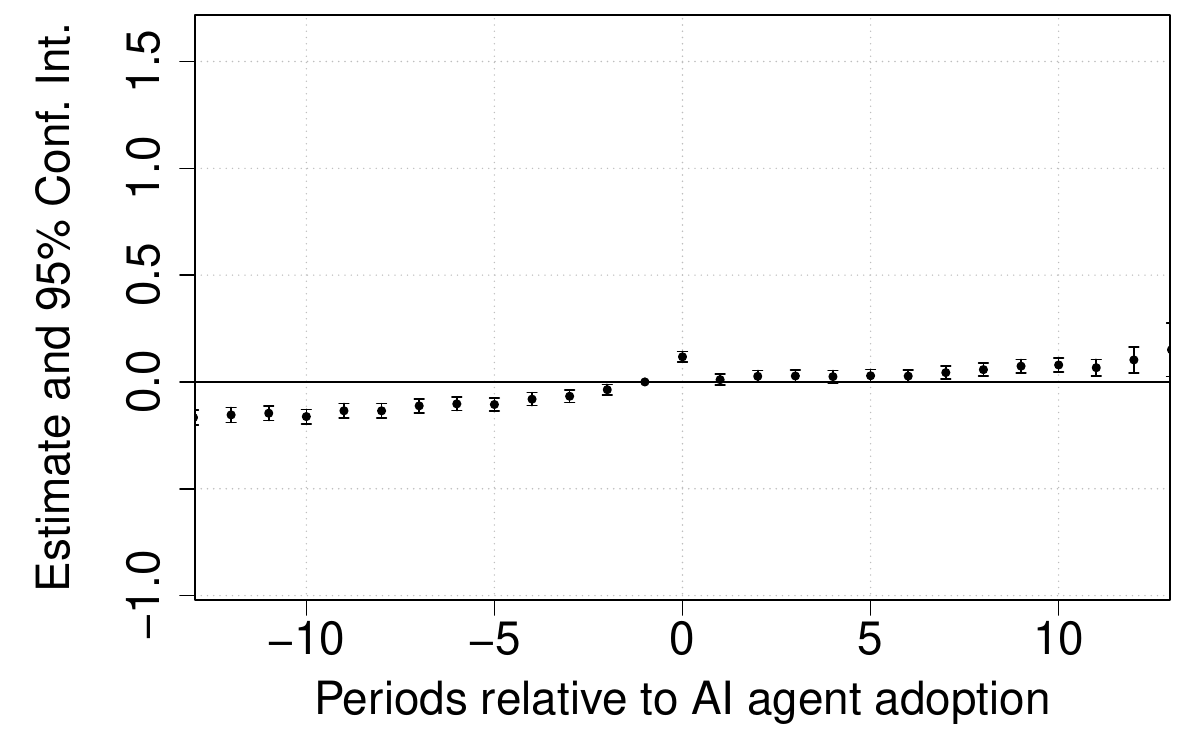}
  \caption{Event study estimates for the log-transformed number of newcomers (secondary outcome of RQ2). Reference period: month $-1$. Error bars denote 95\% confidence intervals.}
  \label{fig:rq2_log}
\end{minipage}
\end{figure}

In the 12 months before adoption, the newcomer ratio showed no detectable difference in trends between the two groups (joint test $p = 0.81$, Figure~\ref{fig:rq2_ratio}); with month $-2$ as the reference month, the coefficient for the month before adoption is also small and not significant ($-0.007$, $p = 0.46$), indicating no significant change ahead of adoption. Therefore, unlike in RQ1, we keep month $-1$ as the reference month. The number of newcomers does not meet this condition: regardless of the reference month, the pre-adoption difference in trends is pronounced ($p < 0.001$, Figure~\ref{fig:rq2_log}). As with the number of contributors in RQ1, it is therefore reported for descriptive purposes only.

Under the main specification (month $-1$ as the reference month, months 1 to 12 after adoption; Section~\ref{sec:causal}), the newcomer ratio decreases by 0.009 on average after agent adoption, about 3\% of the pre-adoption level of the treatment group, and the decrease is not significant ($p = 0.28$; 95\% CI $[-0.024, 0.007]$). With month $-2$ as the reference month, the decrease is 0.015 ($p = 0.064$). When the adoption month is included (months 0 to 12), the estimates are $-0.011$ ($p = 0.14$) and $-0.018$ ($p = 0.025$); with month $-2$, the decrease is therefore significant only when the adoption month is included, and by the decision rule in Section~\ref{sec:causal}, this estimate is considered to depend on the adoption month. In the design that keeps the original matched sample and drops control repositories from the month of their first agent trace onward (Section~\ref{sec:causal}), the estimates are $-0.008$ ($p = 0.27$) with month $-1$ and $-0.012$ ($p = 0.11$) with month $-2$, neither significant. 
On the repositories that were active in at least half of their months 1 to 12 after adoption (Section~\ref{sec:causal}), the estimate is $-0.002$ with either reference month ($p = 0.81$ and $0.82$).
The HonestDiD analysis (Section~\ref{sec:causal}) does not change this picture: with either reference month, the confidence interval of the effect includes zero even if the gap between the two groups remains unchanged after adoption ($[-0.024, 0.007]$ and $[-0.031, 0.001]$), so there is no decrease whose robustness needs to be assessed. 
When projects are required to have at least three or six active months before adoption, the estimate is $0.000$ ($p = 1.00$) and $+0.010$ ($p = 0.22$), respectively. The estimates thus move towards zero as the activity requirement increases, which suggests that any decrease in the full sample is concentrated in projects that were rarely active before adoption.

To examine this, we divided the main matched sample into three groups by the number of months in which the treatment project had human pull request activity before adoption, and estimated the effect separately for each group (Table~\ref{tab:activity_split}). Each treatment repository stays in the same group as its matched controls, and each outcome uses the reference month of its main specification (month $-1$ for the newcomer ratio, month $-2$ for human contributor density). 
Of the 2,246 treatment repositories, 221, 282, and 1,739 fall into the groups with 1--2, 3--5, and at least six active months. 
Four treatment repositories had no active month before adoption and are not assigned to any group, nor are their 7 matched control repositories. Table~\ref{tab:activity_split} reports only the repositories that contribute observations to the estimation: 6 treatment repositories in the first group and 62 of the remaining 3,791 control repositories contribute none, so the first group has 215 treatment repositories.
This grouping was performed after observing the results above and is therefore an exploratory check. 
The decrease is concentrated in the 215 projects that were active in only 1--2 months before adoption ($-0.259$) and, to a lesser extent, in the 282 projects active in 3--5 months ($-0.065$, $p = 0.056$).
The estimates for these two groups are difficult to credit. First, the control projects matched to them are considerably more active: in the 1--2 month group, for example, the median number of active months is 2 for treatment projects and 11 for control projects. Second, the estimates are implausibly large: with month $-2$ as the reference month, the decrease in the 1--2 month group reaches 0.462, more than half of the group's pre-adoption mean.  Moreover, both groups fail the pre-trend test under both reference months ($p < 0.001$ for the 3--5 month group; $p < 0.001$ and $p = 0.021$ for the 1--2 month group). 
The 1–2 and 3–5 month groups therefore do not meet the conditions required for causal inference and are excluded from the substantive conclusion.

In the remaining 1,739 projects that were active in at least six months before adoption (77\% of the treatment group), the newcomer ratio barely changes ($+0.009$, $p = 0.27$), and no difference in pre-adoption trends is detected ($p = 0.97$). For this group, the estimate is fairly precise: the 95\% confidence interval is $[-0.007, 0.025]$, which rules out a decrease of more than 0.7 percentage points. This is consistent with the result of the matching specification defined in advance that requires at least six active months ($+0.010$, $p = 0.22$).
At the same time, with month $-2$ as the reference month, human contributor density from RQ1 decreases by 0.056 in this group, close to 0.048 in the full sample, and the pre-adoption trend test for density is not rejected in this group ($p = 0.11$).
In other words, in the same set of consistently active projects, the number of distinct human contributors behind each pull request decreased, whereas no change was detected in the share of newcomers among human contributors. As for the small, non-significant decrease in the full sample, one possible explanation relates to the stage of the project: in the 1--2 month group, the pre-adoption newcomer ratio averages 0.80 (0.32 in the group active in at least six months), that is, most participants were submitting for the first time; once anyone contributes again, this ratio falls, and such a decline would occur even without agent adoption. Our data do not allow us to test this explanation directly.

\begin{table}[tb]
\centering
\caption{Treatment effects by the number of active pre-adoption months of the treated repository (exploratory analysis). ATT averaged over months 1 to 12 after adoption. Standard errors in parentheses.}
\label{tab:activity_split}
\small
\setlength{\tabcolsep}{4pt}
\begin{tabular}{lrrrrrr}
\toprule
 & \multicolumn{2}{c}{Repositories (T / C)} & \multicolumn{2}{c}{Newcomer ratio (RQ2)} & \multicolumn{2}{c}{HC density (RQ1)} \\
\cmidrule(lr){2-3} \cmidrule(lr){4-5} \cmidrule(lr){6-7}
Active months & Number & Active months & ATT (SE) & Pre-trend $p$ & ATT (SE) & Pre-trend $p$ \\
 & & (median) & & & & \\
\midrule
1--2      & 215 / 404     & 2 / 11  & $-0.259$ (0.047) & $<0.001$ & $-0.004$ (0.042) & 0.019 \\
3--5      & 282 / 508     & 4 / 13  & $-0.065$ (0.034) & $<0.001$ & $-0.018$ (0.032) & 0.127 \\
$\geq 6$  & 1,739 / 2,817 & 23 / 16 & $+0.009$ (0.008) & 0.966    & $-0.056$ (0.008) & 0.114 \\
\bottomrule
\end{tabular}
\end{table}

\begin{tcolorbox}[colback=gray!10, colframe=gray!50, boxrule=0.5pt, left=6pt, right=6pt, top=4pt, bottom=4pt]
\textbf{Answer to RQ2:} After agent adoption, the newcomer ratio shows no detectable change: under the main specification, the estimate is $-0.009$ ($p = 0.28$), and the 95\% confidence interval rules out a decrease of more than 2.4 percentage points. In the matching specification defined in advance that requires at least six active months before adoption, the estimate is +0.010 ($p = 0.22$). Whether the absolute number of newcomers changed cannot be determined, as the parallel trends assumption does not hold for the log-transformed newcomer count.
\end{tcolorbox}

\subsection{RQ3: Agent Pull Requests Recorded as Human Activity}
\label{sec:rq3}

\begin{figure}[tb]
\centering
\begin{minipage}[t]{0.32\textwidth}
  \centering
  \includegraphics[width=\linewidth]{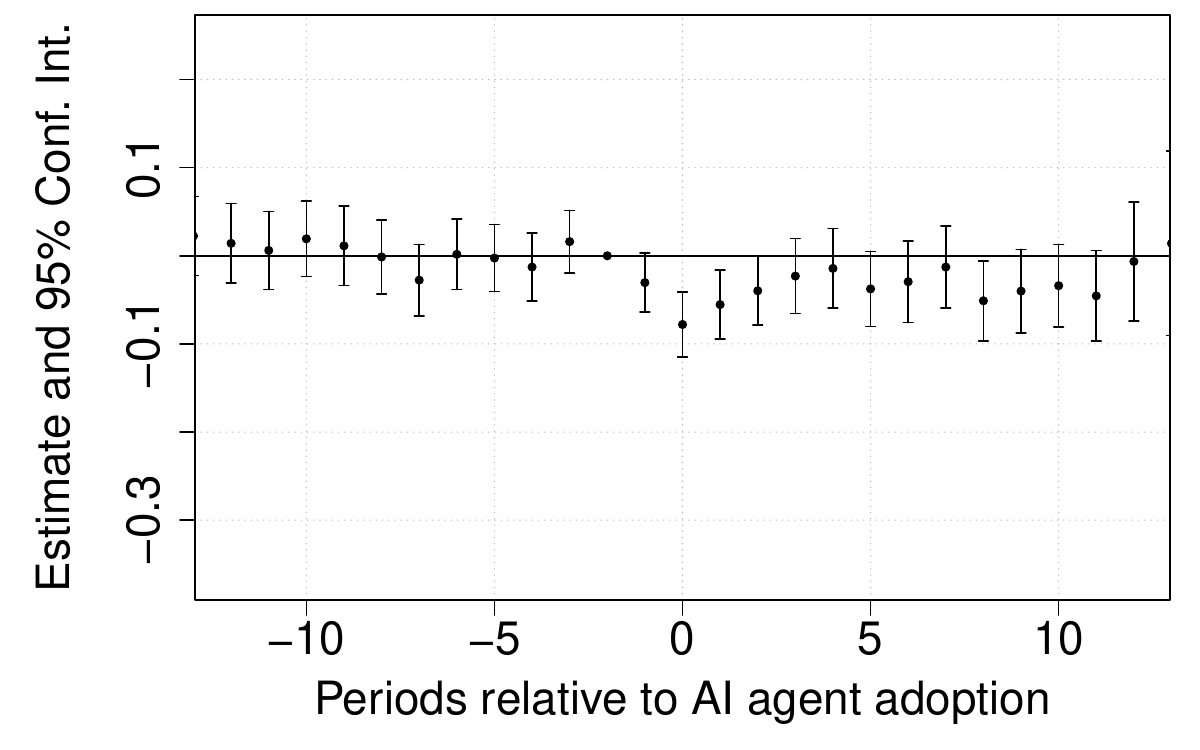}\\
  {\footnotesize (a) Low share (median 2.4\%)}
\end{minipage}\hfill
\begin{minipage}[t]{0.32\textwidth}
  \centering
  \includegraphics[width=\linewidth]{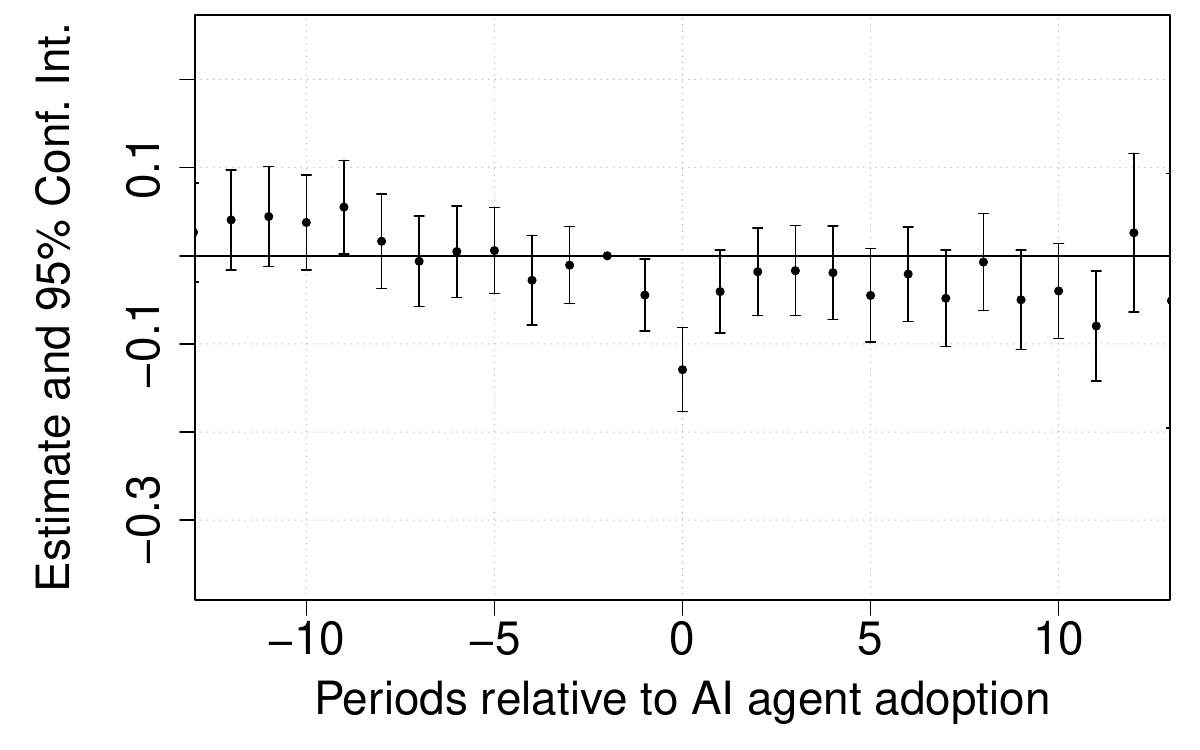}\\
  {\footnotesize (b) Middle share (median 16.6\%)}
\end{minipage}\hfill
\begin{minipage}[t]{0.32\textwidth}
  \centering
  \includegraphics[width=\linewidth]{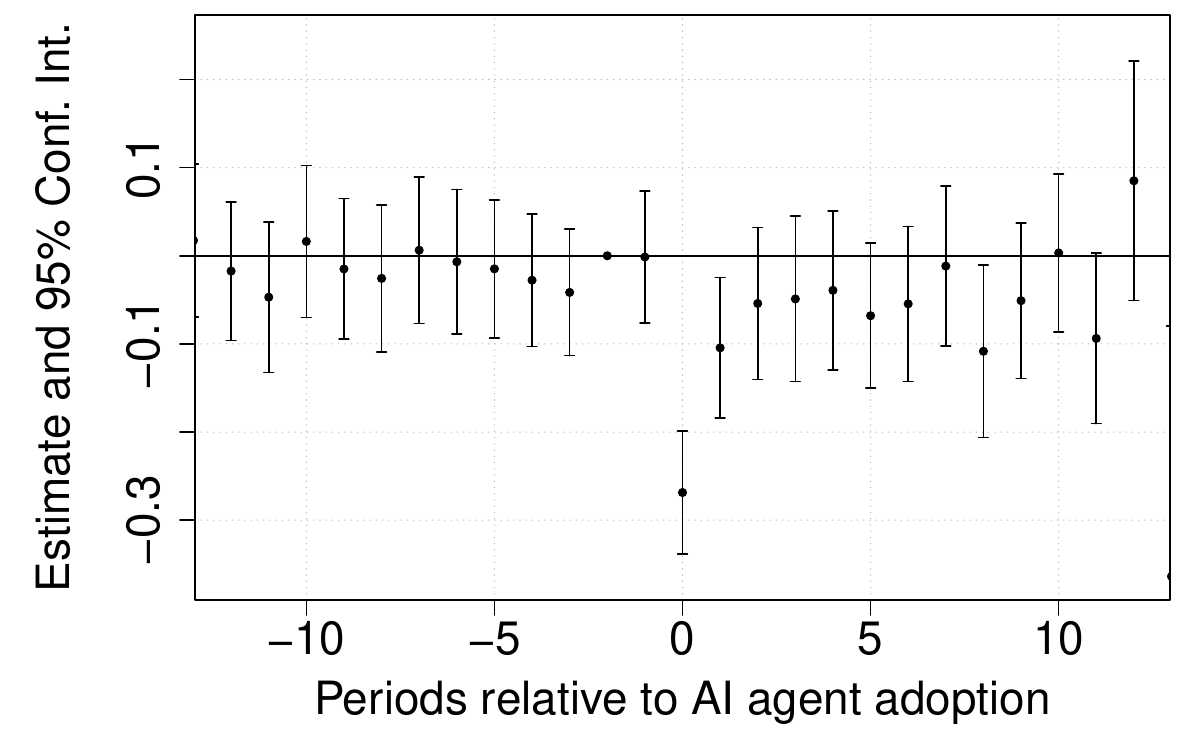}\\
  {\footnotesize (c) High share (median 66.7\%)}
\end{minipage}
\caption{Event study estimates for human contributor density by the share of post-adoption pull requests that are agent pull requests opened by user accounts (RQ3; reference month $-2$). (a) Low, (b) middle, and (c) high share. All panels use the same y-axis.}
\label{fig:rq3}
\end{figure}

To answer RQ3, we compare the change in human contributor density across three groups of repositories with different shares of agent pull requests, as described in Sections~\ref{sec:outcome_var} and~\ref{sec:causal}: Table~\ref{tab:rq3} reports the average effect in each group and the difference between the high and the low group, and Figure~\ref{fig:rq3} shows the monthly changes in each group.

Table~\ref{tab:rq3} divides the treatment repositories into three groups by the share of agent pull requests opened by user accounts among all non-bot pull requests after adoption; the median shares of the three groups are 2.4\%, 16.6\%, and 66.7\%. With month $-2$ as the reference month, the changes in human contributor density in the three groups are $-0.032$, $-0.030$, and $-0.046$, respectively, none of which is significant ($p = 0.067$, $0.15$, and $0.14$). In terms of point estimates, the low and the middle group are almost identical and the high group is slightly larger, but the confidence intervals of the three groups overlap substantially. The difference in the change in density between the high and the low group is $-0.013$, with a 95\% confidence interval of $[-0.083, 0.057]$. With month $-1$ as the reference month, the estimates for the three groups are $+0.001$, $+0.015$, and $-0.052$, none of which is significant either, and the difference between the high and the low group is $-0.053$ (95\% CI $[-0.133, 0.026]$). With month $-2$ as the reference month, the joint pre-adoption test is not significant in any of the three groups; with month $-1$ as the reference month, the test is significant in the middle group ($p = 0.036$). 
This result is therefore inconclusive: the confidence interval of the difference includes zero, but it is also wide enough to include differences as large as the full-sample RQ1 effect (0.048, with month $-2$ as the reference month). It therefore neither shows that the decrease in density is concentrated in projects with a high share of agent pull requests nor rules out this possibility. In addition, the number of pull requests differs greatly across the three groups: the median number of non-bot pull requests after adoption is 149 in the low group, 28 in the middle group, and only 7 in the high group. The comparison between the groups is therefore also a comparison between projects with different numbers of pull requests.

\begin{table}[tb]
\caption{Human contributor density after agent adoption, by the share of post-adoption pull requests that are agent pull requests opened by user accounts (RQ3). ATT averaged over months 1 to 12 after adoption. Standard errors in parentheses.}
\label{tab:rq3}
\centering
\resizebox{\textwidth}{!}{%
\begin{tabular}{lcccccccc}
\toprule
 & \multicolumn{2}{c}{Share of agent PRs} & \multicolumn{2}{c}{Repositories} & \multicolumn{2}{c}{Reference month $-2$} & \multicolumn{2}{c}{Reference month $-1$} \\
\cmidrule(lr){2-3} \cmidrule(lr){4-5} \cmidrule(lr){6-7} \cmidrule(lr){8-9}
Group & Range & Median & Treated & Control & ATT (SE) & Pre-trend $p$ & ATT (SE) & Pre-trend $p$ \\
\midrule
Low    & 0.0--6.9\%  & 2.4\%  & 337 & 535 & $-0.032$ (0.018) & 0.404 & $+0.001$ (0.018) & 0.191 \\
Middle & 7.0--33.3\% & 16.6\% & 340 & 579 & $-0.030$ (0.021) & 0.300 & $+0.015$ (0.020) & 0.036 \\
High   & 34.5--100\% & 66.7\% & 333 & 619 & $-0.046$ (0.031) & 0.860 & $-0.052$ (0.037) & 0.912 \\
\midrule
\multicolumn{5}{l}{High minus low (95\% CI)} & \multicolumn{2}{l}{$-0.013$ [$-0.083$, $0.057$]} & \multicolumn{2}{l}{$-0.053$ [$-0.133$, $0.026$]} \\
\bottomrule
\end{tabular}%
}
\end{table}

Figure~\ref{fig:rq3} shows, separately for the three groups, the monthly estimates of human contributor density in the 12 months before and after adoption, each expressed relative to the second month before adoption; the three panels share the same vertical axis. Before adoption, the estimates for the low and the high group are close to zero and show no clear trend; the middle group is slightly above zero in the earliest months and then gradually declines to around zero, and the joint test for months $-12$ to $-3$ is not significant ($p = 0.300$). In the month before adoption, the estimate for the middle group is slightly below zero (about $-0.04$). All three groups show a clear decrease in the adoption month, and the point estimate increases with the share of agent pull requests: about $-0.08$ in the low group, $-0.13$ in the middle group, and $-0.27$ in the high group; as described in Section~\ref{sec:causal}, this month is not included in the ATT. In months 1 to 12 after adoption, the estimates of the three groups lie in a similar range, mostly between $-0.1$ and $0$, and in most months the confidence interval includes zero. The high group fluctuates more and has wider confidence intervals, at about $-0.11$ in months 1 and 8 and close to zero in month 10. Months 11 and 12 can only be estimated from repositories that adopted early, and their confidence intervals are wide, so they should be interpreted with caution. In other words, judging from the figure, the most visible difference between the three groups occurs in the adoption month; after that month, the changes in density in the three groups lie in a similar range, and in most months the confidence interval includes zero.

To understand whether the change in density comes from the number of pull requests or from the number of people, we use the same model to estimate the changes in $\log(\mathit{pr\_count} + 1)$ and $\log(\mathit{human\_contributors} + 1)$ in each of the three groups. In the low and the middle group, the joint pre-adoption test is significant for most of these estimates, so we do not interpret them. In the high group, the pre-adoption test is not significant for either variable; with month $-2$ as the reference month, the number of pull requests increases by about 8\% after adoption ($p = 0.059$), whereas the number of human contributors remains almost unchanged ($+0.010$, $p = 0.69$), and the results are similar with month $-1$ as the reference month. This pattern is consistent in direction with an increase in pull requests while the number of people submitting them stays the same, but because the decrease in density in the high group is itself not significant and this decomposition is only descriptive, we draw no further conclusion from it.

\begin{tcolorbox}[colback=gray!10, colframe=gray!50, boxrule=0.5pt, left=6pt, right=6pt, top=4pt, bottom=4pt]
\textbf{Answer to RQ3:} 
Inconclusive. The change in human contributor density after adoption does not differ detectably between repositories with high and low shares of agent pull requests recorded under developer accounts ($-0.013$, 95\% CI [$-0.083, 0.057$]), but the interval is too wide to rule out a difference as large as the full-sample effect. Repositories with a high share also have far fewer pull requests after adoption than those with a low share.
\end{tcolorbox}

\section{Discussion}
\label{sec:discussion}

\subsection{Implications for Maintainers and Platforms}
\label{sec:implications}
\paragraph{Newcomer participation.}
A common concern is that AI coding agents may crowd out newcomers' opportunities to participate in open-source projects. 
Contributions by AI coding agents consist predominantly of features and fixes~\cite{robbes2026agentic}, and one of the main barriers newcomers face when joining open-source projects is finding a suitable task to start with~\cite{steinmacher2015systematic}. If tasks that could otherwise serve as starting points are increasingly completed by agents, the contribution opportunities available to newcomers may shrink.
Our results offer a partial answer to this concern. 
In the full sample, human contributor density decreased by 0.048 after agent adoption (month $-2$ as the reference month), 
about five fewer distinct people behind every 100 pull requests, whereas no change was detected in the share of newcomers among human contributors ($-0.009$, $p = 0.28$), and the estimate rules out a decrease of more than 2.4 percentage points (Sections~\ref{sec:humanpartici} and~\ref{sec:newcomer}). The same pattern appears in the exploratory analysis of projects that were consistently active before adoption (77\% of the treatment group): density decreased by 0.056, whereas the newcomer ratio barely changed ($+0.009$).
The difference between the two measures comes from their denominators: 
the denominator of density is the number of pull requests opened by non-bot accounts in a given month,
whereas the denominator of the newcomer ratio includes only human contributors. Taken together, this pattern suggests that the change lies in the volume of pull requests, that is, each human contributor corresponds to more pull requests, rather than in fewer newcomers among the humans who participate. 
Whether the additional pull requests were submitted by agents on their own or by the same people with the help of agents cannot be distinguished with our data (Section~\ref{sec:threats}).
Regarding newcomers, Xu et al. use the first commit of an agent configuration file as the adoption signal, likewise find no evidence of crowding out in newcomer inflow, the handling of first contributions, or newcomer retention, and observe an increase in newly opened good first issues after adoption (which the authors treat as correlational evidence)~\cite{xu2026decoupling}. The two studies differ in their adoption definitions, samples, and measures, and neither finds evidence that agent adoption reduces newcomer participation in the measures it examines; in our study, this applies to the share of newcomers, as the change in their absolute number could not be determined (Section~\ref{sec:newcomer}). However, neither study measures who actually completes entry-level tasks. Projects that rely on good first issues to attract newcomers may still want to watch whether these tasks are increasingly assigned to agents.

\paragraph{Community health metrics.}
Metrics that use the number of pull requests or the number of participants per pull request to gauge project activity and community health~\cite{goggins2021open} may lose comparability over time once a project adopts agents: an increase in pull requests no longer implies that more people are participating. Earlier research has shown that bot accounts distort measures such as team size and project activity~\cite{dey2020detecting}. Agent pull requests submitted under developer accounts break the same assumption that one account corresponds to one person (Section~\ref{sec:huamn_in_osp}), and they are harder to detect than bot accounts. 
Recent work has argued that agent-generated traces violate the human-origin assumption underlying trace-based software measurement~\cite{bock2026ground}, and the CHAOSS community has begun building tooling to track disclosed AI contributions. To the best of our knowledge, however, no study has yet empirically quantified how agent contributions affect community health metrics.

\paragraph{Platform labeling.}
Because agent pull requests are submitted either from bot accounts or from developers' own accounts (Section~\ref{sec:2.1}), and projects have not converged on a convention for disclosing AI use~\cite{hoffmann2024generative}, repository records alone cannot reliably separate human from agent contributions. 
Ghaleb~\cite{ghaleb2026fingerprinting} further notes that agent pull requests submitted from developers' accounts put the ``human'' portion of datasets such as AIDev at risk of mislabeling. Our measurement is subject to the same limitation (Section~\ref{sec:threats}). Beyond post hoc identification (Section~\ref{sec:3.4}), a consistent and queryable label for agent-generated pull requests provided by platforms would help maintainers distinguish human from agent contributions more clearly, and help researchers assess the effects of agents on open-source communities more accurately.

\subsection{Lessons Learned: Pitfalls in Estimating the Effects of Agent Adoption}

In the course of our analysis, we revised several of our initial design choices, relied on one check planned in advance, and replaced our main data source; these lessons apply equally to work that uses similar data to study the effects of agent adoption.

When we first constructed the control group, we estimated propensity scores and performed matching using six aggregate measures of project size and activity (the number of users involved, pull requests, forks, comments, issues, and total events). This approach had two problems. First, these measures capture only the level of activity, not its trend, whereas difference-in-differences requires precisely that the two groups follow parallel trends before adoption; differences in levels are removed by differencing, but differences in trends are not. Second, the measures for the treatment group covered only the pre-adoption period, whereas those for the candidate control group covered the entire observation period, so the two groups were measured inconsistently. As a result, the matched sample showed persistent differences in trends across the adoption point in the event study plots. We then re-matched based on the monthly panel, using the same pre-adoption time window for both groups and adding the pre-adoption trends of the number of human contributors, pull requests, and newcomers as covariates. 
The standardized mean differences of these trend covariates were 0.428 to 0.437 before matching and fell below 0.03 after matching (Table~\ref{tab:balance}). This lesson shows that assessing matching quality cannot rest on balance in size alone; pre-adoption trends also need to be included in the matching and their balance reported. 

A second issue concerns the newcomer ratio. In the main specification, the newcomer ratio decreased slightly after adoption ($-0.009$, $p = 0.28$), but the estimate moved towards zero when the matching required projects to have been active for at least three or six months before adoption ($0.000$ and $+0.010$). Grouping by the number of active months showed that the decrease came almost entirely from projects that had been active in only one or two months before adoption ($-0.259$). The newcomer ratio of these projects is inherently unstable, and the control projects matched to them were considerably more active; the matching was balanced overall but not within this subsample (Section~\ref{sec:newcomer}). 
An earlier version of this analysis, based on GHArchive data and posted as a preprint~\cite{zhang2026augmentation}, found a significant decrease in the full sample and reported it as a reduction in the newcomer share; there, too, the decrease came almost entirely from these projects. With outcomes rebuilt from the GitHub API (Section~\ref{sec:datacollection}), the full-sample decrease is no longer significant, and we no longer draw that conclusion.
This lesson shows that, when using ratio outcomes, one should check whether the estimates are driven by projects with very low activity; inconsistency across robustness specifications is itself a diagnostic signal worth investigating.

A third issue concerns the timing of adoption. Adoption dates derived from repository records capture only agent use that leaves a record, so they can lag behind actual adoption, and the month immediately before the recorded adoption may already be a post-adoption month for some repositories. If this month serves as the reference period, a decrease in it shows up as a violation of parallel trends. We encountered this pattern in an earlier version of our analysis; we then adopted a protocol that reports results under both month $-1$ and month $-2$ and specifies when month $-2$ is treated as primary (Section~\ref{sec:causal}), and applied it to all estimates reported here. For human contributor density, the protocol selected month $-2$ (Section~\ref{sec:humanpartici}). This lesson shows that, when adoption timing comes from records that capture only observable agent use, one should expect it to lag behind actual adoption and fix in advance both the reference months to be reported and the rule for choosing between them.

A fourth issue concerns the data source. Compared with the GitHub API, GHArchive records about 86\% of pull requests between January 2023 and April 2025, but only 32\% in October 2025 and 6.5\% in May 2026 (Section~\ref{sec:datacollection}). A similar decline has been reported independently for star events in the GHArchive repository, beginning around June 2025 and becoming much sharper from February 2026~\cite{gharchive2026issue}, which suggests that the problem is not specific to pull request events or to our data pipeline. Because the missing records affect pull requests more than contributors, ratio outcomes such as human contributor density are distorted in exactly the period in which most repositories adopted agents. We found this problem only while examining a different measurement question, after the main analysis had been completed. We therefore rebuilt all outcome variables from the API and re-estimated all results. This lesson shows that GHArchive coverage should be checked against the GitHub API before GHArchive is used to measure activity after mid-2025, especially for ratio outcomes.

\subsection{Threats to Validity}
\label{sec:threats}
\paragraph{Construct Validity}
We use pull request submissions as a proxy for human contributor activity. Contributors who submit commits directly without opening pull requests are not captured in our measurements, which may lead to an underestimation of human contributor counts. 
However, pull-request-based contribution workflows are the dominant practice in modern open-source projects~\cite{gousios2014exploratory}, and this limitation applies equally to treatment and control repositories; it is therefore unlikely to produce systematic differences between the two groups, unless agent adoption itself changes the share of contributions made through pull requests.
Human contributor density is defined as the ratio of human contributors to 
pull requests in a given repository-month. Our bot filtering rule is also applied to the pull request count, so agent pull requests submitted from bot accounts are not counted in the denominator, whereas agent pull requests submitted from developers' own accounts are not only counted in the denominator, but their submitters are also counted as human contributors~\cite{ghaleb2026fingerprinting}. In other words, this measure captures the number of distinct human accounts behind each pull request. Its decrease after adoption may stem either from agents taking over part of the pull requests or from the same developers submitting more pull requests with the help of agents, and does not necessarily mean that human participation decreased. 
A decrease could also arise if pull request submission became concentrated among fewer contributors for reasons unrelated to agents that coincide with adoption; the difference-in-differences design rules this out only to the extent that the parallel trends assumption holds.
Because agent pull requests that carry no signature cannot be told apart from human pull requests, we cannot distinguish between these two cases; RQ3, which uses the agent pull requests that do carry a signature, could not settle this question either (Section~\ref{sec:rq3}).

Our bot filtering excludes accounts that the GitHub API reports as bots, as well as accounts whose login names contain ``bot'' or that are one of three commonly used automation accounts (Section~\ref{sec:datacollection}). Two residual threats remain. First, the name rule is case-sensitive, so user accounts that contain ``Bot'' only in upper case and are not reported as bots by the API are not excluded. Second, automation accounts that are registered as ordinary user accounts and do not follow these naming conventions may be incorrectly counted as human contributors.

We extended the historical window for newcomer identification back to January 2020 to mitigate left-truncation bias. However, for contributors who had contribution history in a repository prior to January 2020, our data may still incorrectly classify them as newcomers, potentially leading to a slight overestimation of newcomer counts.

Our treatment onset is the earlier of two months: the month of each repository’s first agent-authored pull request in AIDev, and the month of its first agent trace in our scan (Section~\ref{sec:dataset}). Both sources capture only agent use that leaves a record, such as a pull request in AIDev’s collection or a default signature; agent use outside AIDev’s coverage, or with signatures removed, is not detected. Our treatment onset can therefore be later than, but not earlier than, the true adoption date; we discuss the consequences for the reference month under internal validity. Moreover, our treatment captures only agents that leave a record on GitHub.
Moreover, our treatment captures only agents that leave a record on GitHub. AI assistance through chat interfaces or IDE code completion leaves no such record and is not observed in either group. Our estimates therefore compare projects that adopted autonomous agents with projects that, as far as we can observe, did not, rather than projects with and without any AI assistance.

\paragraph{Internal Validity}
We constructed the control group through propensity score matching, balancing both the pre-adoption level of activity and its trend;  
After matching, the SMD of every covariate is below 0.07 in the matched sample (Table~\ref{tab:balance}) and below 0.08 in the final sample (Section~\ref{sec:datacollection}).
However, matching can only balance observed covariates. If unobserved factors affect both the adoption of AI agents and the outcome variables, the estimates may still be biased. 
In addition, the matching covariates are computed up to April 2025, whereas most treatment repositories adopted agents in June or July 2025, so for these repositories the one or two months immediately before adoption are not covered by the matching. These months correspond to months $-1$ and $-2$, whose event study coefficients are reported in Section~\ref{sec:humanpartici} and Section~\ref{sec:newcomer}; factors that change only around the adoption month, however, cannot be ruled out by matching or by the pre-adoption tests.
The diagnostics and sensitivity analyses described in Section~\ref{sec:causal} indicate that our main estimates tolerate only small violations of parallel trends (Section~\ref{sec:results}); we therefore interpret them as suggestive evidence rather than definitive causal effects.
Because the recorded onset can lag behind actual adoption, the pre-adoption period of some repositories may contain post-adoption months. Following the protocol in Section~\ref{sec:causal}, we therefore treat month $-2$ as the primary reference month for human contributor density. The decrease in month $-1$ that triggered the protocol is consistent with adoption beginning before the recorded month, but it may also reflect a genuine divergence immediately before adoption, and our data cannot distinguish the two. Under the second reading, month $-1$ is the appropriate reference; the estimate is then smaller ($-0.019$) and tolerates smaller violations of parallel trends (Section~\ref{sec:humanpartici}). We therefore report the effect as a range from $-0.019$ to $-0.048$ rather than as a single estimate.
The stratification of the newcomer ratio by the number of active pre-adoption months was performed after observing the full-sample result and is explicitly exploratory. Its result for consistently active projects ($+0.009$) agrees with the robustness matching specification defined in advance that requires at least six active months ($+0.010$; Section~\ref{sec:newcomer}). Both ratio outcomes are defined only for months with human pull request activity, and adopting projects were more likely than control projects to remain active after adoption; our estimates therefore apply to active months. 
Restricting the sample to repositories that were active in at least half of their months 1 to 12 after adoption leaves the primary estimates nearly unchanged ($-0.045$ for density and $-0.002$ for the newcomer ratio), although the density estimate with month $-1$ as the reference month is no longer significant (Sections~\ref{sec:humanpartici} and~\ref{sec:newcomer}).
Because this restriction itself selects repositories on post-adoption activity, it does not remove a possible bias from the difference in activity between the two groups; the ratio estimates should therefore be read as effects within active repository-months, and we regard this as a limitation of the estimand rather than a resolved threat.

One unobserved factor that may affect both adoption and the outcomes is repository visibility; prior research shows that project popularity is closely associated with the onboarding of newcomers~\cite{fronchetti2019attracts}. However, our data do not contain a stars measure that is consistent across the two groups: for treatment repositories, stars is the total star count at the time of AIDev data collection, whereas for candidate control repositories, it is the number of stars gained during the observation period. Stars were therefore not included in the matching. By directly balancing the pre-adoption level and trend of newcomer counts in the matching, we partly mitigate the bias arising from differences in visibility, but cannot fully rule it out. Moreover, stars have limitations as a measure of visibility: developers star repositories for diverse reasons, and star counts may also be influenced by promotion and dissemination strategies~\cite{borges2018s}.

The control group may be contaminated. We excluded repositories that appear in AIDev from the control group (Section~\ref{sec:control}), but control repositories may still have adopted agents after the AIDev data collection ended, or may have submitted agent-generated pull requests from developers' own accounts throughout, and thus not have been recorded by AIDev. Given the already high adoption rate of agents on GitHub~\cite{robbes2026agentic}, such cases may not be rare. If the control group is partly treated and agent use affects control repositories in the same direction, the difference between the two groups shrinks, so our estimates would understate rather than overstate the effects of agent adoption.
Our scan for agent traces (Section~\ref{sec:screening}) confirms this concern: at least 31.8\% of candidate control repositories showed agent use, most of them after June 2025.
We address this concern in two ways. First, the agent-trace screening described in Section~\ref{sec:screening} removed all 2,554 candidate control repositories with a detected trace from the pool before matching. Second, in the robustness design described in Section~\ref{sec:causal}, we retain the original matched sample but use each control repository only up to the month before its first detected trace; the estimates remain similar (ATT $= -0.044$ with month $-2$ as the reference month, compared with $-0.048$ in the main analysis), suggesting that residual contamination does not materially affect the conclusions.

\paragraph{External Validity}
Our treatment group consists of repositories with more than 100 GitHub stars, which excludes smaller projects with lower visibility. The findings of this study may not generalize to repositories below this threshold, as the dynamics of AI agent adoption and its effects on contributor ecosystems may differ in smaller or less active projects.
Moreover, the treatment group is dominated by OpenAI Codex and GitHub Copilot (Table~\ref{tab:agent_repos}), while agents differ considerably in how they work and how their contributions are submitted; we did not estimate effects separately by agent, so our conclusions may not apply to agents that are less represented in the sample. Adoption is also concentrated between May and July 2025; given how quickly agent capabilities and usage practices evolve, our conclusions reflect this period.

\subsection{Future Work}
Future research could advance in several directions.

\paragraph{Newcomer outcomes and long-term participation}
As noted in Section~\ref{sec:implications}, neither study measures who completes entry-level tasks.
Comparing the task-type distributions of agent contributions and newcomers’ first contributions (e.g., by tracking whether good first issues are ultimately completed by agents or by newcomers) would allow a more direct test of whether agents systematically occupy newcomers’ entry paths.
Even when the relative share of newcomers remains stable, their subsequent outcomes may have already changed: the merge rate of newcomers’ first pull requests, and whether they continue to contribute in the following months, are both important indicators for assessing the impact of agent adoption on community health; constructing contributor-level longitudinal panels would allow these questions to be answered directly.
Furthermore, our observation window does not allow longer-term outcomes such as newcomer retention or growth into core contributors to be assessed; longer observation periods, or identification strategies that do not rely on parallel trends such as instrumental variables, would help address this limitation.

\emph{Review roles and burden.} This study does not examine review work as an outcome; review activity enters only as a matching covariate (Section~\ref{sec:second_stage}). An earlier version of this paper~\cite{zhang2026augmentation} reported an increase in review depth based on GHArchive data. Because GHArchive’s coverage of pull request events declines after mid-2025 (Section~\ref{sec:datacollection}), similar declines have been reported for other event types~\cite{gharchive2026issue}, and we did not retrieve review events through the GitHub API, that result could not be re-examined with the data used here, and we do not carry the analysis over to the present version. How agent adoption affects review depth, review participation rates, and the distribution of review burden across contributors remains open. In addition, the degree of overlap between pull request submitters and reviewers, and whether agent adoption affects these two roles differently, remain unexamined. Contributor-level longitudinal panels would allow changes in participation on the submission and review sides, as well as the redistribution of review burden across contributors, to be tracked separately.

\paragraph{Attribution and measurement}
Our data cannot separate agent-mediated pull requests from human pull requests (Section~\ref{sec:threats}). Bringing identification methods based on behavioral features~\cite{ghaleb2026fingerprinting, ghaleb2026agentag} into repository-level analyses could separate human activity from agent-mediated activity, and thus allow a more accurate interpretation of changes in human contributor density. 
Relatedly, we plan to examine in follow-up work the comparability of community health metrics based on the number of pull requests and participants before and after agent adoption, and to explore possible corrections.

\paragraph{Agent heterogeneity}

This study treats the adoption of all five agents as a single treatment (Section~\ref{sec:threats}).
Estimating effects separately by agent would help reveal which types of agents have the greatest impact on human contributor ecosystems. As more repositories enter a stable post-adoption period, revisiting RQ3 in a larger sample, in which the difference between groups can be estimated more precisely, and separately by agent type, would be a valuable direction.

\section{Conclusion}
\label{sec:conclusion}
This paper presents a large-scale empirical study of how human contributor ecosystems in open-source software change after AI coding agent adoption. Using a staggered difference-in-differences design with agent-trace screening of candidate control repositories, two-stage propensity score matching, and HonestDiD sensitivity analysis, we address three questions: how the structure of human participation changes after agent adoption, whether the relative participation share of newcomers decreases, and whether the decline in contributor density is associated with agent pull requests recorded under developer accounts. The screening itself showed that at least 31.8\% of candidate control repositories outside AIDev carried traces of agent use. The results are consistent with a pattern of \emph{augmentation with dilution}: human contributor density declines by 0.019 to 0.048 after adoption, depending on the reference month, while no change is detected in the relative participation share of newcomers, so that each human contributor corresponds to a larger number of pull requests. Whether this decline reflects agent pull requests recorded under developer accounts remains inconclusive. Our data do not show that agents displace human participants. As agents continue to spread across open-source ecosystems, the evolution of human participation structures warrants sustained attention from the research community.

Future research can advance along three directions: tracking at the task level who ultimately completes good first issues, to test whether agents substantively occupy newcomers' entry paths; constructing contributor-level longitudinal panels to assess newcomer retention and growth into core contributors; and examining how agent adoption affects review depth, review participation, and the distribution of review burden across contributors.

\bibliographystyle{ACM-Reference-Format}
\bibliography{main}

\end{document}